\documentclass{article}

\usepackage{arxiv}
\usepackage{amsmath,amssymb,amsfonts}
\usepackage{array}
\usepackage{multirow}
\usepackage[utf8]{inputenc} 
\usepackage[T1]{fontenc}    
\usepackage{hyperref}       
\usepackage{url}            
\usepackage{booktabs}       
\usepackage{nicefrac}       
\usepackage{microtype}      
\usepackage{graphicx}
\usepackage{float}
\usepackage{doi}
\usepackage{xcolor, soul} 
\usepackage{blindtext}
\usepackage[numbers]{natbib}
\usepackage{relsize}
\usepackage{caption}
\usepackage{subcaption}
\usepackage{subfloat}

\title{Decoding Neuronal Networks: A Reservoir Computing Approach for Predicting Connectivity and Functionality}
\author{ Ilya Auslender \\
	Department of Physics\\
	University of Trento\\
	Via Sommarive 14, 38123, Trento, TN, Italy \\
 \texttt{ilya.auslender@unitn.it}
	\And
	Giorgio Letti \\
    Centre for Integrative Biology (CIBIO)\\
    University of Trento\\
    Via Sommarive 9, 38123 Trento, TN, Italy
\And
Yasman Heydari\\
Center for Mind/Brain Sciences (CIMeC)\\
University of Trento\\
Corso Bettini, 31, 38068 Rovereto, TN, Italy
\And
Clara Zaccaria\\
Department of Physics\\
University of Trento\\
Via Sommarive 14, 38123, Trento, TN, Italy \\
 \And
	Lorenzo Pavesi \\
	Department of Physics\\
	University of Trento\\
	Via Sommarive 14, 38123, Trento, TN, Italy
}

\date{}

\renewcommand{\shorttitle}{}

\hypersetup{
pdftitle={A template for the arxiv style},
pdfsubject={q-bio.NC, q-bio.QM},
pdfauthor={David S.~Hippocampus, Elias D.~Striatum},
pdfkeywords={First keyword, Second keyword, More},
}

\begin{document}
\maketitle

\begin{abstract}
In this study, we address the challenge of analyzing electrophysiological measurements in neuronal networks. Our computational model, based on the Reservoir Computing Network (RCN) architecture, deciphers spatio-temporal data obtained from electrophysiological measurements of neuronal cultures. By reconstructing the network structure on a macroscopic scale, we reveal the connectivity between neuronal units. Notably, our model outperforms common methods like Cross-Correlation and Transfer-Entropy in predicting the network’s connectivity map. Furthermore, we experimentally validate its ability to forecast network responses to specific inputs, including localized optogenetic stimuli.
\end{abstract}


\section{Introduction}
In the field of neuroscience, electrophysiological studies offer a wide-ranging perspective on the intricate interplay among different cell types at varying scales \cite{llinas1988intrinsic}. These investigations span from understanding the function of individual cells to unraveling the dynamics within complex systems composed of numerous cells, \cite{contreras2004electrophysiological}, all with the aim of capturing a holistic view of brain activity. As biological systems grow in complexity, the challenge intensifies when attempting to analyze or model their behavior. To address this, a plethora of models has emerged, ranging from single-cell descriptions (such as the Hodgkin–Huxley model \cite{hodgkin1952quantitative}) to large-scale population models \cite{marder2011multiple,gerstner2002spiking}. These models explore diverse aspects, including biophysical properties (e.g., membrane voltage) and information propagation (e.g., spike trains). Some innovative approaches leverage experimental data to adapt computational models that mirror biological systems \cite{natschlager2003computer,yada2021physical,dockendorf2009liquid}. Machine learning and deep learning techniques play a pivotal role in training these models to achieve desired outcomes. While such approaches may be computationally intensive for certain research questions, they offer a practical avenue for constructing computational tools across various applications.

In this work, we propose an artificial neural network model for interpreting electrophysiological signals within intricate networks. Our approach uncovers connectivity patterns among different sampled regions in the network and captures dynamic interactions between them. The model is based on the \textit{Reservoir Computer Network} (RCN) architecture \cite{lukovsevivcius2009reservoir}, leveraging information gleaned from electrophysiological measurements in complex neural circuits. These circuits defy easy comprehension through standard measurement analyses, leading us to model them as nonlinear networks with internal random connections. Notably, our model outperforms common methods, including Cross-Correlation (CC) \cite{perkel1967neuronal,salinas2001correlated}  and Transfer-Entropy (TE) \cite{PhysRevLett.85.461}, in accurately predicting the network’s connectivity map (CM). Furthermore, we demonstrate the model’s ability to forecast the spatio-temporal response of a given network to specific inputs. To validate our predictions, we combine experimental measurements using microelectrode arrays (MEA) with \textit{in-vitro} mice cortical neurons \cite{auslender2023integrated} and numerical simulations conducted using the NEST simulator \cite{Gewaltig:NEST}.

\section{Reservoir Computing Model for Electrophysiological Signals Decoding}
\begin{figure}
    \centering
    \subfloat[]{\includegraphics[width=0.2\textwidth]{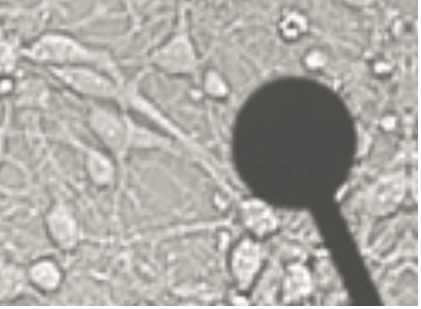}
    \label{fig: electrode & neurons}}
    \subfloat[]{\includegraphics[width = 0.55\textwidth]{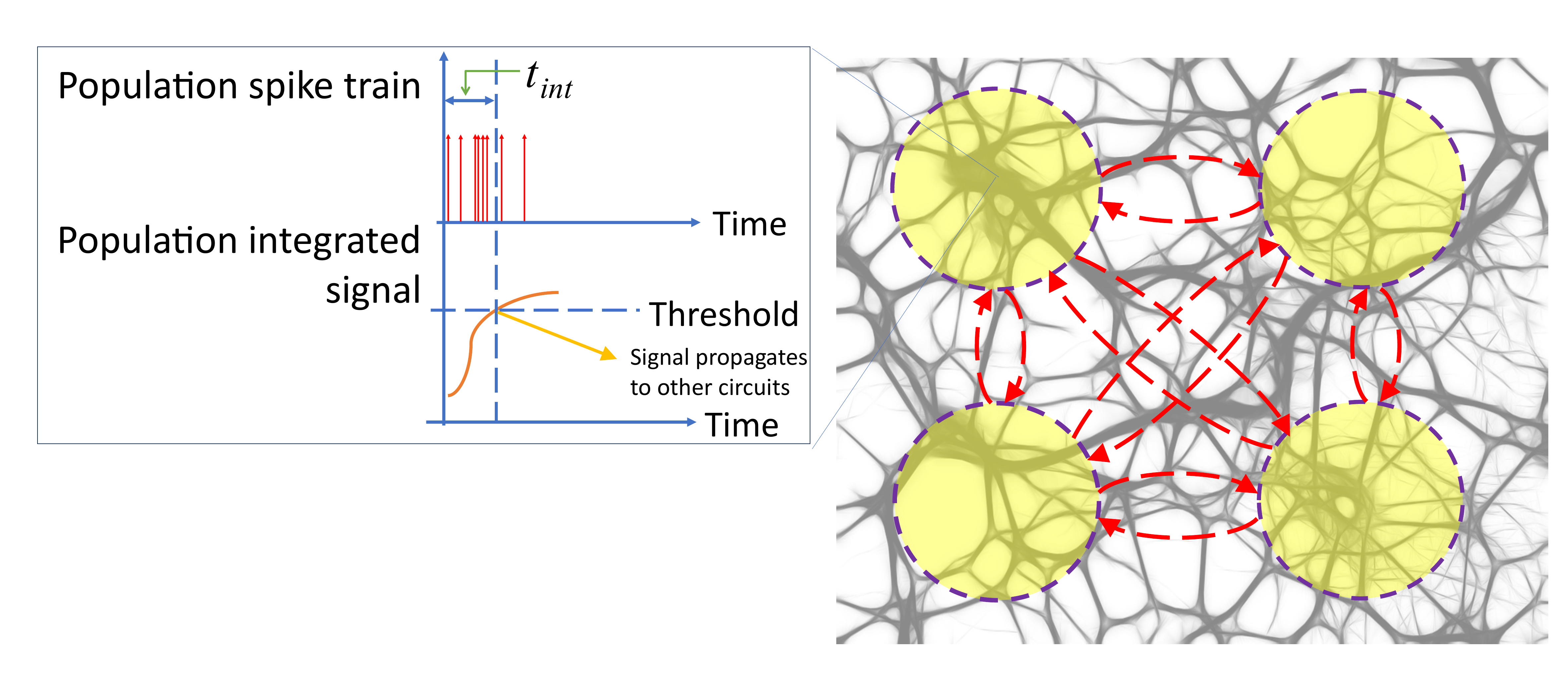}
    \label{fig: integration of circuits}}
    \hfill
    \subfloat[]{\includegraphics[width=0.8\textwidth]{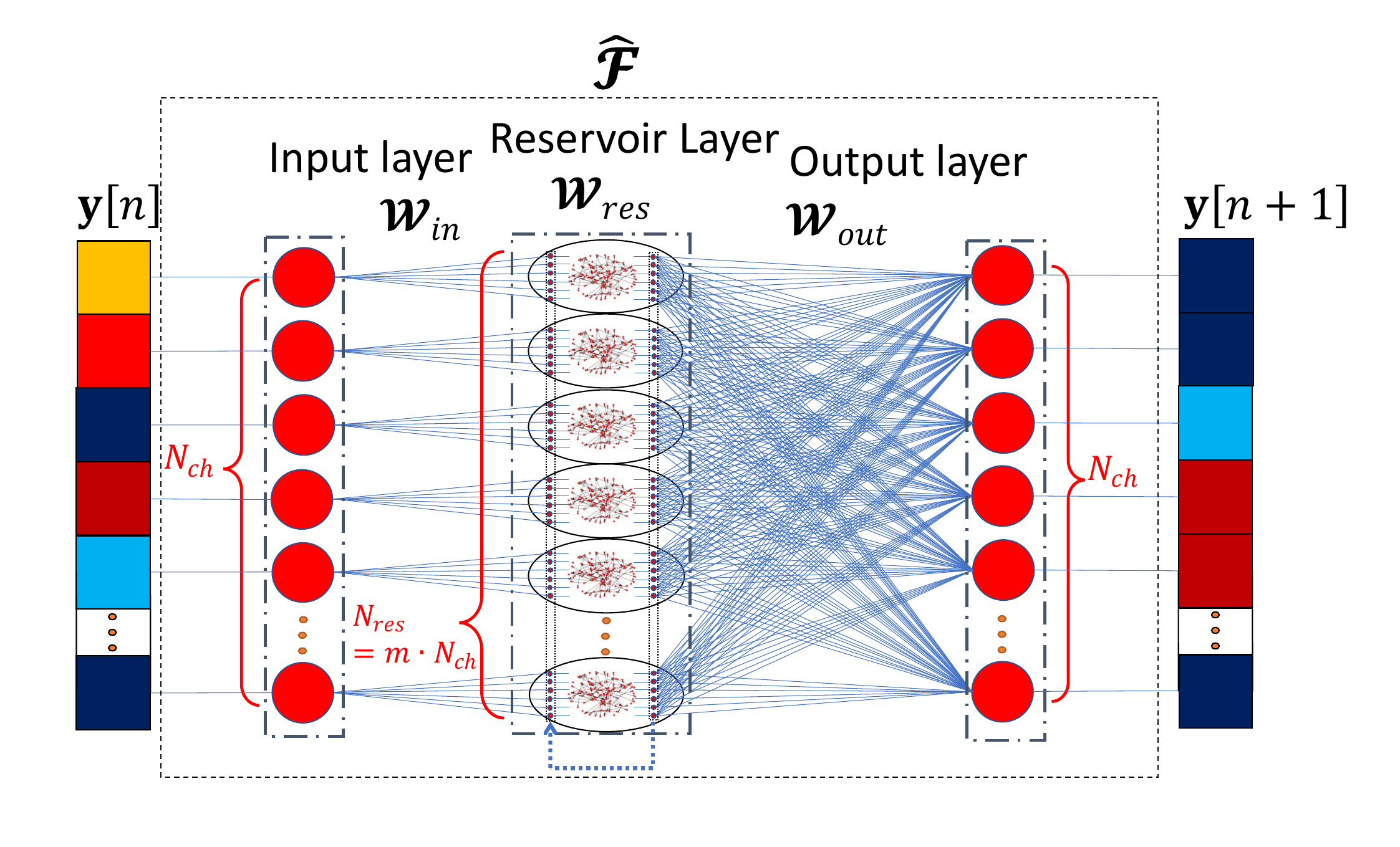}
    \label{fig: artificial neural network}}
    \hfill
    \subfloat[]{\includegraphics[width = 0.5\textwidth]{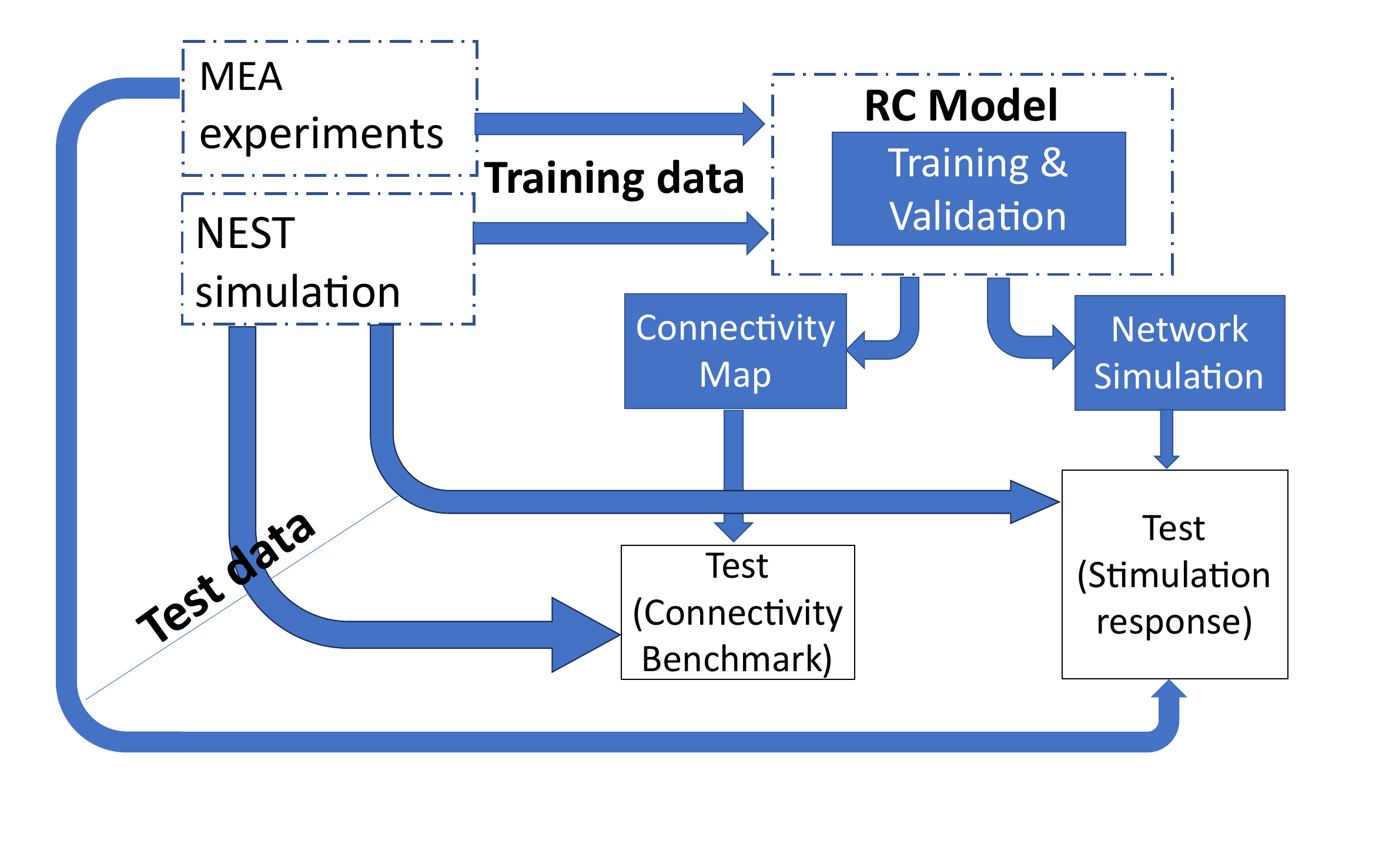}\label{fig: process}}
    \caption{Description of the RC model. (a) A microscope image of amicroelectrode (diameter of 30 $\mu$m) surrounded by cultured neurons. This image illustrates how each electrophysiological measurement site samples signals from a complex circuit, whose morphology is not resolvable by simple methods. (b) An illustrative image of how the signals are intercepted by each electrode (yellow circles) from the biological neuronal network in the background. The spike trains collected at each electrode are then forming integrated signals which propagate within the network. (c) The artificial neural network (ANN) design of the RC model. It consists of three layers (input, reservoir, output) and a recurrent branch. It describes how the state of the network $\mathbf{y}$ (described by $N_{ch}$ samples of the measured signals from each electrode) is processed from time step $n$ to time step $n+1$. (d) A block diagram illustrating the sequential computational processes, including training, validation, and testing, conducted in this study.}
    \label{fig: introduction}
\end{figure}

The Reservoir Computing (RC) architecture models electrophysiological data obtained from biological neuronal networks using multi-electrode arrays (MEA). The objective of the RC model is to decode these signals, ultimately reconstructing a network structure within the measured neuronal culture. In a  MEA measurement, each electrode (channel) captures signals from a nearby neuronal population (microcircuit), as depicted in Fig. \ref{fig: electrode & neurons}. In our artificial neural network (ANN), each such microcircuit or population is treated as a node of the macroscopic network (representing the measurement domain), and their interconnections are to be reconstructed during the training phase. This approach enables the RC model to unveil the intricate connectivity and relationships within the biological neuronal networks. 

To describe the ANN, let us consider a discrete time domain denoted by $n=1,2,3...$ corresponding to real time units ($t$) by the relation:
\begin{equation}
\label{eq: discrete time}
    t = n\cdot t_{int} \;.
\end{equation}
Since each neuronal microcircuit can be likened to a recurrent spiking neural network with a leaky integrator \cite{gerstner2002spiking} (as shown in Fig. \ref{fig: integration of circuits}), $t_{int}$ is defined as the integration time that characterizes the time duration of which a neuronal microcircuit processes incoming information. Next, we assume that the relevant node state corresponds to the firing rate measured by each MEA channel. Let us represent the network node states using a vector $\mathbf{y}[n]\in \mathbb{R}^{N_{ch}\times 1}$, where $\mathbf{y}[n]$ represents the instantaneous spike rate (ISR) measured by each of the MEA electrodes at time step $n$ (with $N_{ch}$ being the number of measurement channels). To predict the network's next state $\mathbf{y}[n+1]$ based on the current and past states  $\mathbf{y}[n],\mathbf{y}[n-1],\mathbf{y}[n-2]\ldots$, we have designed the ANN illustrated in Fig. \ref{fig: artificial neural network}. In this architecture, the input layer is connected to a layer of $N_{ch}$ independent reservoirs, each representing the neuronal microcircuits probed by the MEA electrodes. The reservoir has an internal dimension $m$. Additionally, each reservoir is connected to all the nodes via a trainable interconnecting matrix at the output layer.
The state evolution from $\Tilde{\mathbf{y}}[n]$ to $\Tilde{\mathbf{y}}[n+1]$ (the tilde sign indicates predicted states) through the ANN is described by the following set of equations:

\begin{subequations}
\label{eq: dynamics}
\begin{align}
\mathbf{x}_{in}[n] = \mathcal{W}_{in}\Tilde{\mathbf{y}}[n]
\label{eq: input}\\
\mathbf{x}[n] = \mathbf{f}_{NL} \Bigl(\hat{\mathcal{S}}\cdot\left( \mathbf{x}_{in}[n]+\alpha\mathcal{W}_{res}\mathbf{x}[n-1] \right)\Bigr)
\label{eq: reservoir equation non-linear}\\
\Tilde{\mathbf{y}}[n+1] = \mathcal{W}_{out}\mathbf{x}[n] + \mathbf{b}
\label{eq: output}
\end{align}
\end{subequations}

where:
\begin{itemize}
    \item[] $\mathbf{x}_{in},\mathbf{x}\in \mathbb{R} ^ {N_{res}\times 1}$ are the reservoir input and actual state respectively ($N_{res} = m \times N_{ch}$ is the reservoir layer dimension).
    \item[] $\mathcal{W}_{in}\in \mathbb{R} ^ {N_{res}\times N_{ch}},\hat{\mathcal{S}}\in \mathbb{R} ^ {N_{res}\times N_{res}},\mathcal{W}_{res} \in \mathbb{R} ^ {N_{res}\times N_{res}},\mathcal{W}_{out} \in \mathbb{R} ^ {N_{ch}\times N_{res}}$ are input, synaptic, reservoir and output matrices respectively.
    \item[] $0 < \alpha < 1$ is the reservoir layer internal memory strength.
    \item[] $\mathbf{b}\in \mathbb{R} ^ {N_{ch}\times 1}$ is a bias vector.
    \item[] $\mathbf{f}_{NL}(\mathbf{x}): \mathbb{R} ^ {N_{res}\times 1} \rightarrow \mathbb{R} ^ {N_{res}\times 1}$ is a vector valued nonlinear activation function. 
\end{itemize}

Therefore,
\begin{equation}
    \Tilde{\mathbf{y}}[n+1] = \hat{\mathcal{F}} \Bigl\{\Tilde{\mathbf{y}}[n]\Bigr\} \; .
    \label{eq: full}
\end{equation}

To train the RC model, we utilized both experimental and simulated ISR sequences. Given that the ANN belongs to the RCN type, training is exclusively applied to the output layer (Eq. \eqref{eq: output}) using a linear regression. This leverages the intricate dynamics embedded in the reservoir layer (described by Eqs. \eqref{eq: input},\eqref{eq: reservoir equation non-linear}).
Let us consider the training data matrix $\mathbf{Y}\in \mathbb{R}^{N_{ch}\times N_t}$, which governs $N_t$ time steps of the observed network state. In other words, $\mathbf{Y} = \Bigl[\mathbf{y}[1],\mathbf{y}[2]\ldots \mathbf{y}[N_t]\Bigr]$. The training task involves optimizing the matrix of weights $\mathcal{W}_{out}$ and the biases $\mathbf{b}$ of the output layer, in order to minimize the error between $\mathbf{Y}$ and $\Tilde{\mathbf{Y}}$. We use the Lasso regression method \cite{tibshirani1996regression}, employing the following loss function:

\begin{subequations}
\label{eq: lasso regression}
    \begin{align}
        \mathcal{L} = \frac{1}{N_{ch}} \sum_{i=1} ^{N_{ch}} \left \{ \left\langle \left\|\Tilde{y}_i[n] - y_i[n] \right\|^2\right\rangle_n  +\lambda\sum _j \left |W_{out}^{i,j}\right|\right \}\\
        \Tilde{y}_i[n] = \sum_j W_{out} ^{i,j} \cdot x_j[n]+ b_i \; .
    \end{align}
\end{subequations}

Here $\lambda$ represents the lasso regularization parameter, and the $\langle \cdot \rangle _n$ notation denotes averaging along the time steps dimension. Then the objective of the regression is expressed as:
\begin{equation}
    \left\{\mathcal{W}_{out},\mathbf{b}\right\} = \underset{{\mathbf{w},\mathbf{b}}}{\arg\min} \bigl(\mathcal{L}\bigr)
    \label{eq: objective}
\end{equation}
where $\mathbf{w}$ encompasses all the weights in the matrix $\mathcal{W}_{out}$. The choice of the lasso regression method stems from its effectiveness in identifying essential weights (preventing them from exploding) and efficiently excluding unnecessary weights by setting them to zero.

\section{Results}
\subsection{Retrieval of Connectivity Map}
\label{sec: connectivity results}
Here, we demonstrate the first feature of our Reservoir Computing (RC) model: the ability to derive the connectivity map between measurement sites from the spatio-temporal dynamics encoded in electrophysiological signals. Connectivity refers to the weighted relationships between different nodes of the network.

Let us assume that the Artificial Neural Network (ANN) has successfully learned the dynamics of a given biological network (i.e., achieved low loss values during training). The dynamics described by Eq.\eqref{eq: dynamics} show the evolution of the network state $\mathbf{y}[n]$, which occurs on an intrinsic network composed of fundamental connections with varying strengths. We hypothesize that these connections can be associated with the linear regime of the dynamics. Therefore, we represent the nonlinear dynamics described by Eq.\eqref{eq: full}, which governs the evolution of the network state $\mathbf{y}[n]$, as follows:

\begin{equation}
\label{eq: basic DE}
    \mathbf{y}[n+1] = \hat{\mathcal{F}} \Bigl\{\mathbf{y}[n]\Bigr\} =\mathcal{T}_0\mathbf{y}[n]+\mathcal{G}\Bigl\{\mathbf{y}[n],\mathbf{y}[n-1],\mathbf{y}[n-2] \ldots\Bigr\}
\end{equation}
where $\mathcal{T}_0$ is defined as the \textit{Intrinsic connectivity matrix} (ICM), describing the linear relation between consecutive states. $\mathcal{G}$ denotes a nonlinear transformation that captures higher-order interactions, embodying the memory of the network. By linearizing Eq.\eqref{eq: dynamics}, we obtain:
\begin{equation}
    \mathcal{T}_0 = \mathcal{W}_{out}\hat{\mathcal{S}}\mathcal{W}_{in}
    \label{eq: intrinsic conn matrix}
\end{equation}
This allows us to depict the morphology of the network using a directed graph of nodes and edges. Each node represents a neuronal population, whereas the edges represent the connections, with their appointed weights. The weight of a connection signifies the coupling strength between the two populations.

\begin{figure}
    \centering
    \begin{subfigure}[]{0.7\textwidth}
        \includegraphics[width=\textwidth]{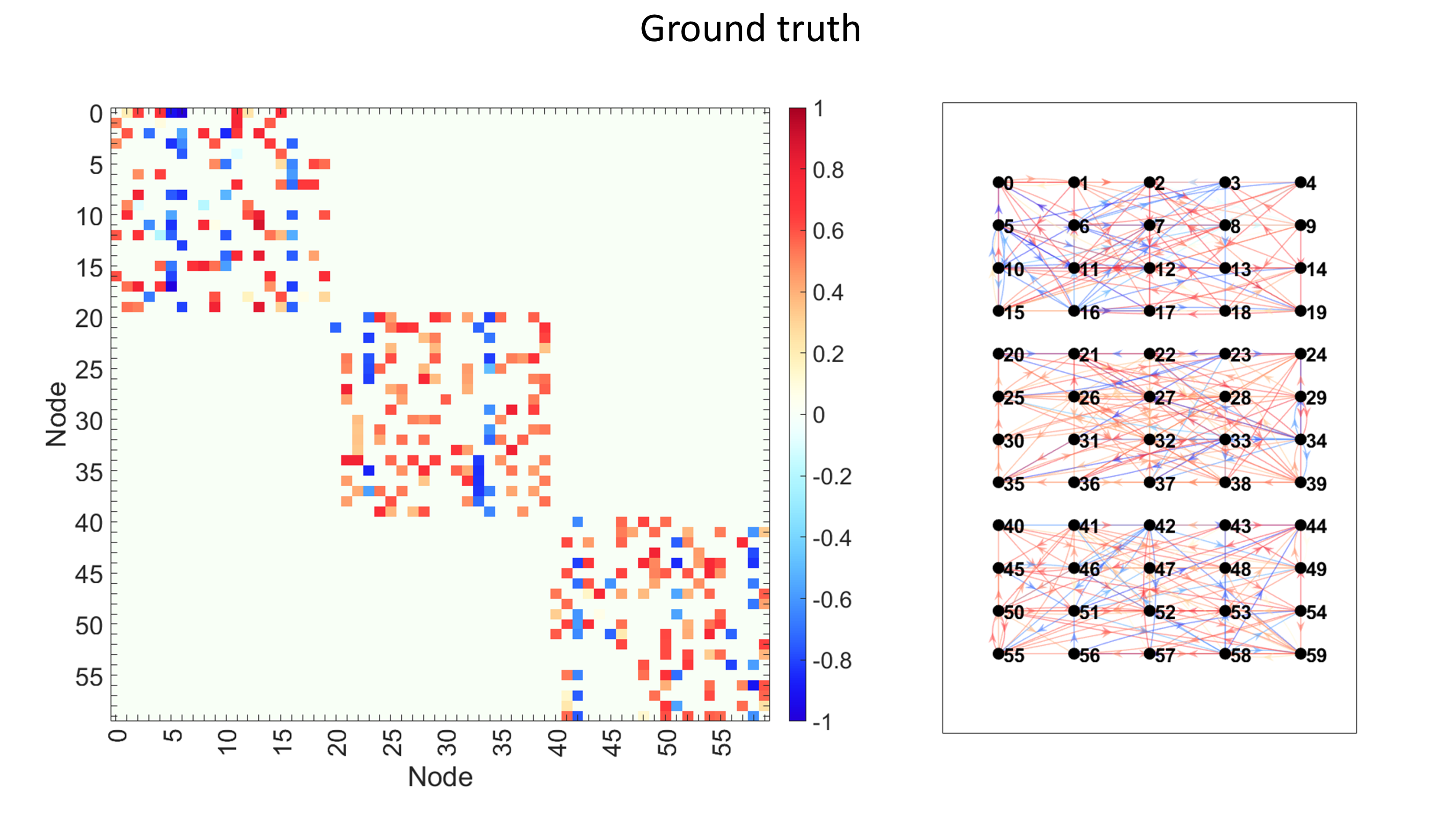}
        \caption{}
        \label{fig: GT CM}
    \end{subfigure}
    \begin{subfigure}[]{0.7\textwidth}
        \includegraphics[width=\textwidth]{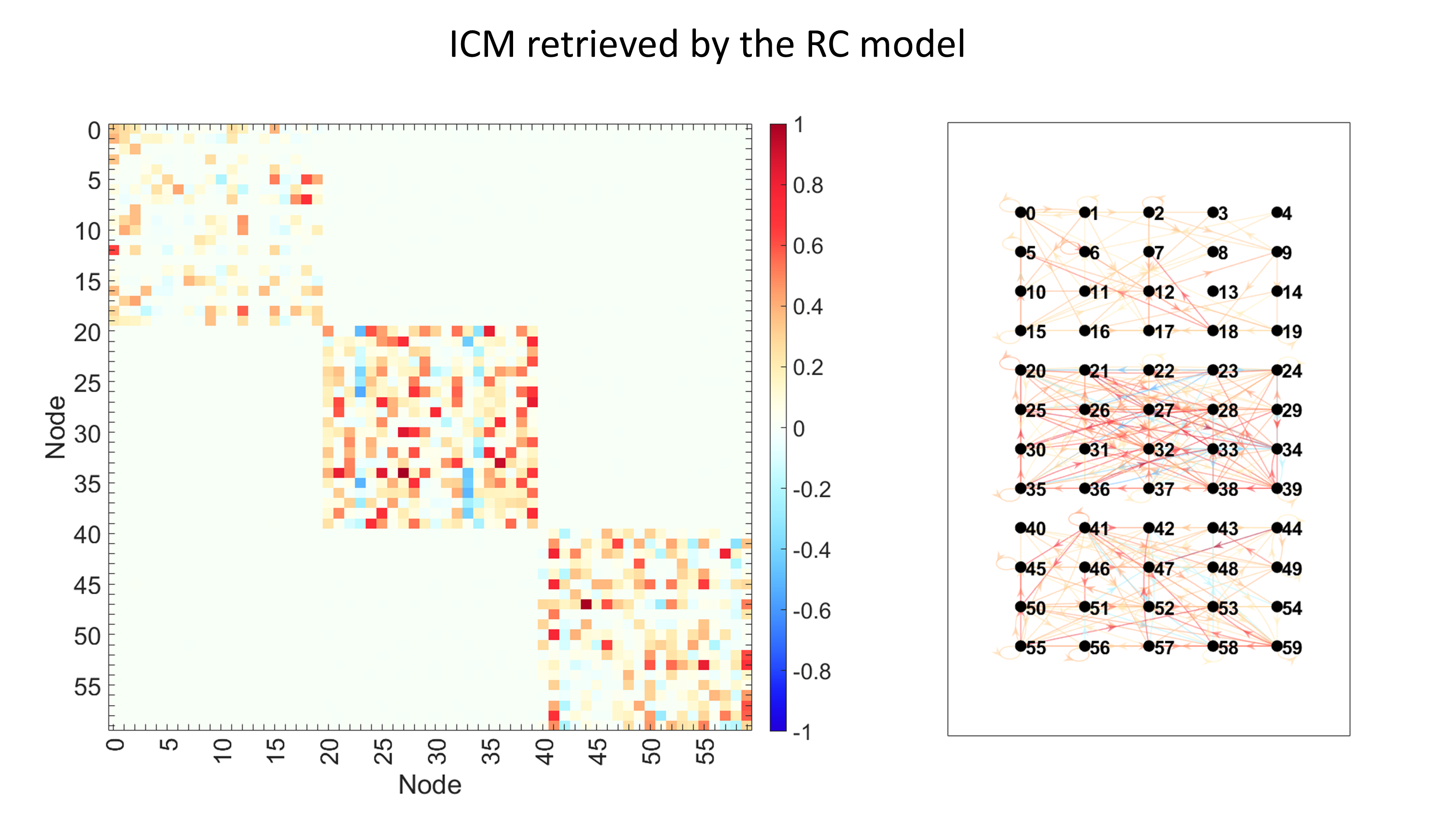}
        \caption{}
        \label{fig: ICM}
    \end{subfigure}
    \begin{subfigure}[]{0.3\textwidth}
        \includegraphics[width = \textwidth]{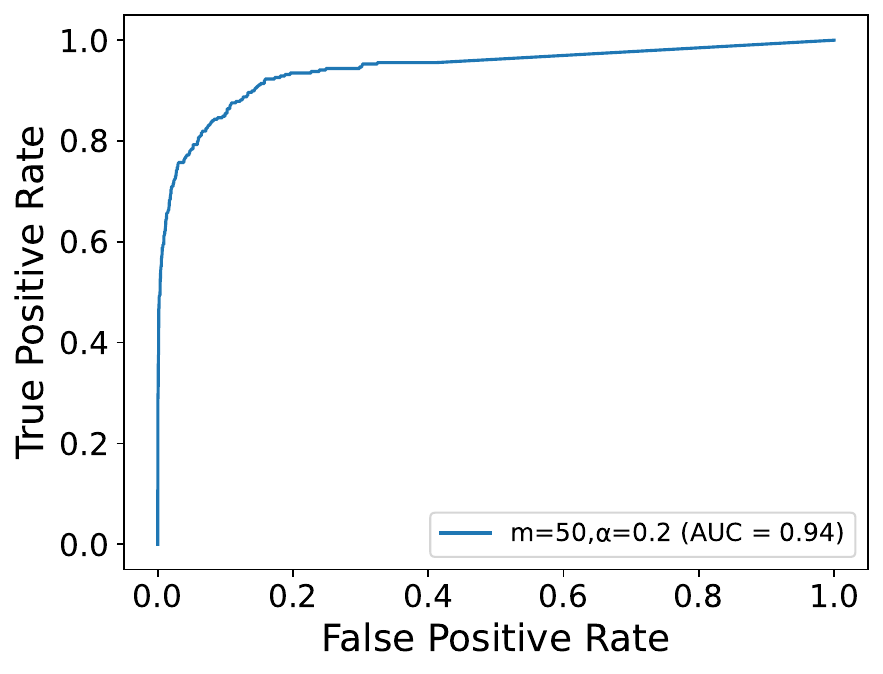}
        \caption{}
        \label{fig: ROC CM}
    \end{subfigure}
    \caption{Connectivity map obtained by the RC model trained on NEST simulation data. (a), (b) Connectivity matrix (left) and connectivity graph map (right) for both the ground-truth network (a) and the the RC model (b). Color bar refers to the connection strengths (weights). The ground-truth network was simulated by a NEST simulator, producing an electrophysiological activity which was then decoded by the RC model obtaining the \textit{Intrinsic Connectivity Matrix} (ICM). The nodes in the graphs represent neuronal population (or circuit) as probed by the electrodes (channels) and the edges represent their connections. Connection strengths in the RC model are deduced by the weight matrix $\mathcal{T}_0$ (Eq. \ref{eq: intrinsic conn matrix}). Each weight was normalized by the highest element in absolute value and color coded between -1 and 1. Note that the RC model predicted with a very high accuracy the existence of 3 uncoupled groups of nodes. In addition, the model distinguished with a high accuracy between excitatory connections (positive weights) and inhibitory (negative weights) yielding a Person correlation $\rho = 0.72$. The connectivity map in (b) displays only weights with absolute values exceeding 0.2. (c) ROC curve to which corresponds a $AUC = 0.94$. In this example, the RC model had a memory strength $\alpha = 0.5$ and a reservoir dimension $m = 50$.}
    \label{fig: CM example}
\end{figure}

To validate our hypothesis, we utilize the ICM to represent the actual connectivity between the nodes of the network, with each node corresponding to a neuronal population. As a benchmark, we construct a simulation of a neuronal network with ground-truth connections between groups of neurons (named ground-truth network). This network is built and simulated using the NEST simulator \cite{Gewaltig:NEST}. In the simulation, we create a virtual array of electrodes that sample signals from a defined population of neurons. This setup mirrors conditions found in experimental electrophysiological MEA recordings. Next, we benchmark the RC model with both the ground-truth network and experimental data. We evaluate prediction accuracy using two distinct methods:
\begin{enumerate}
    \item Receiver Operating Characteristic (ROC) Curve Analysis \cite{macmillan2004detection,fawcett2006introduction}. This method illustrates the model’s predictive performance in a binary context, distinguishing between the presence and absence of connections among the nodes. We assess the performance using the Area Under the Curve (AUC) of the ROC curve, where a value of 1 means that all the connections are properly detected by the RC model.
    \item Pearson Correlation ($\rho$) with ground-truth \cite{WeissteinCorrelation,pearson1895notes}. This metric provides a quantitative measure of the model's prediction quality. It not only captures the presence or absence of connections among the nodes but also quantifies the accuracy in predicting connection weights. These weights may encompass both excitatory (positive values) and inhibitory (negative values) connections within the network.
\end{enumerate} 

\begin{figure}[t!]
    \centering
    \includegraphics[width=0.8\textwidth]{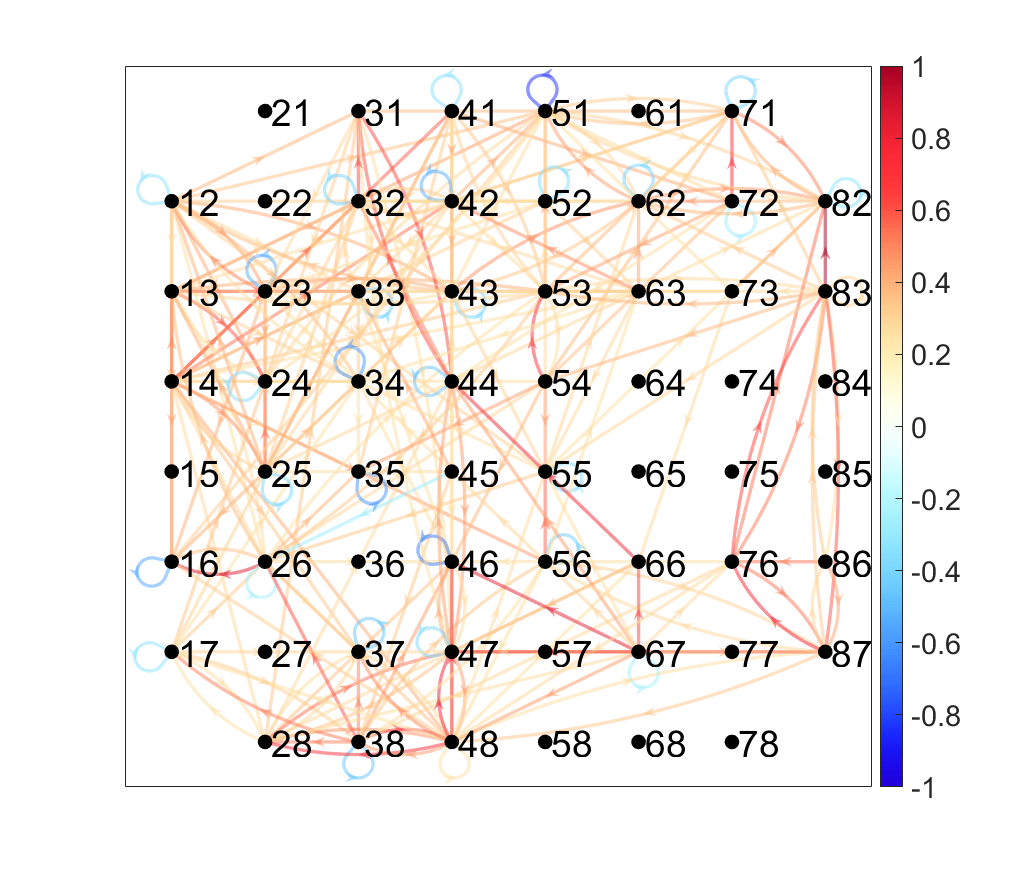}
    \caption{Connectivity map derived by the RC model from experimental data collected using a 60-electrode MEA and a mice cortical neuronal culture. The connectivity weights are normalized based on the highest absolute value within the obtained ICM model. The color code refers to the color bar on the right, scaled between -1 and 1. To enhance clarity, connections with weights below $|0.25|$ have been omitted.}
    \label{fig:CM exp}
\end{figure}

To effectively elucidate the model's capacity, we showcase its performance by simulating the MEA signals from a neuronal population formed by three distinct, non-interconnected groups of nodes (node=neurons population probed by a single electrode). Each group of 20 nodes (out of a total of 60) possesses internal connections, while inter-group connections are absent. Figure \ref{fig: GT CM} shows the input connectivity matrix or connectivity graph map used to simulate the data used to train the RC model. After training, by linearizing the RC model we extracted the the connectivity map shown in Fig. \ref{fig: ICM} which is compared to the ground-truth maps (Fig. \ref{fig: GT CM}). The obtained accuracy results of $AUC = 0.94$ (Fig. \ref{fig: ROC CM}) and $\rho = 0.72$ (Pearson correlation) demonstrate the capacity of the RC model in predicting the connectivity. It not only distinguishes the existence and absence of connections ($AUC$) but also retrieves the weight values and distinguishes between excitatory and inhibitory connections (positive or negative sign) ($\rho$). In this way we can use MEA recordings to decipher intricate connectivity patterns by the analysis of spatio-temporal signals from an actual neuronal culture. An example of a connectivity map obtained by the RC model from experimental MEA data can be found in Fig. \ref{fig:CM exp}.

\begin{figure}[t!]
    \centering
    \begin{subfigure}[b]{0.43\textwidth}
        \includegraphics[width=\textwidth]{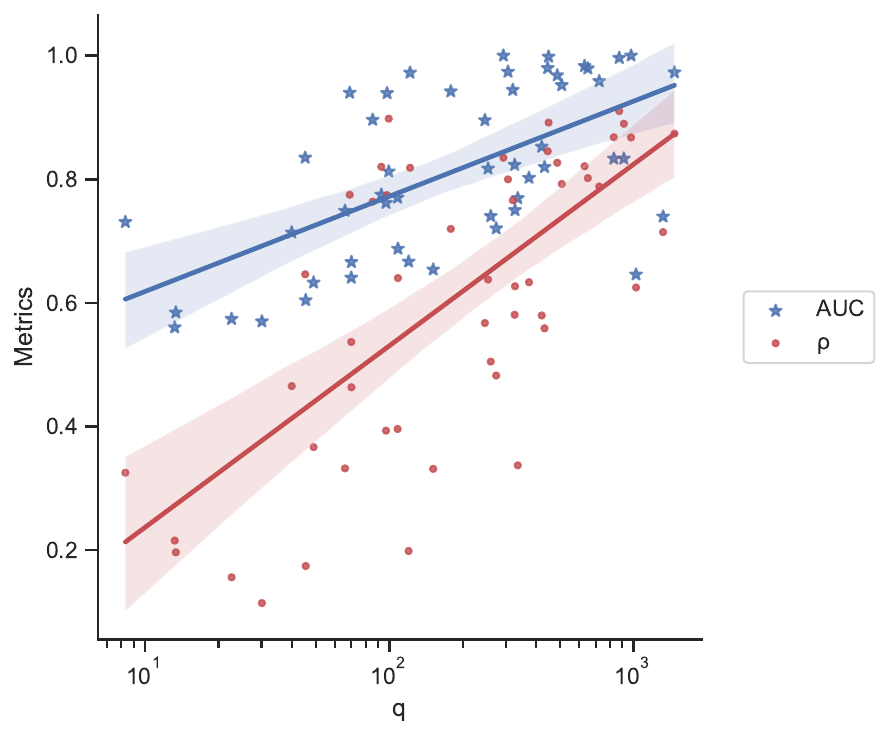}
        \caption{}
        \label{fig:connectivity auc & rho vs q}
    \end{subfigure}
    \begin{subfigure}[b]{0.55\textwidth}
        \includegraphics[width=\textwidth]{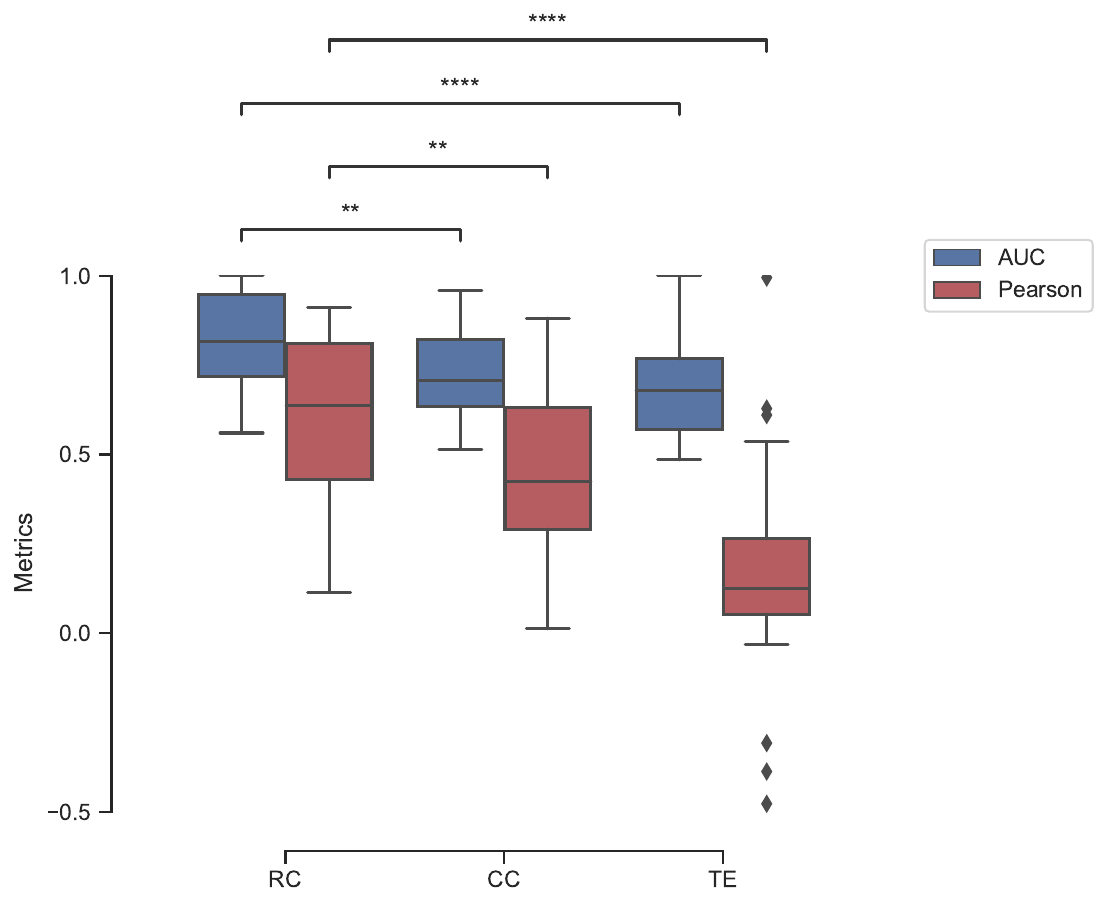}
        \caption{}
        \label{fig:connectivity general, algorithms}
    \end{subfigure}
    \hfill
    \begin{subfigure}[b]{0.48\textwidth}
        \includegraphics[width=\textwidth]{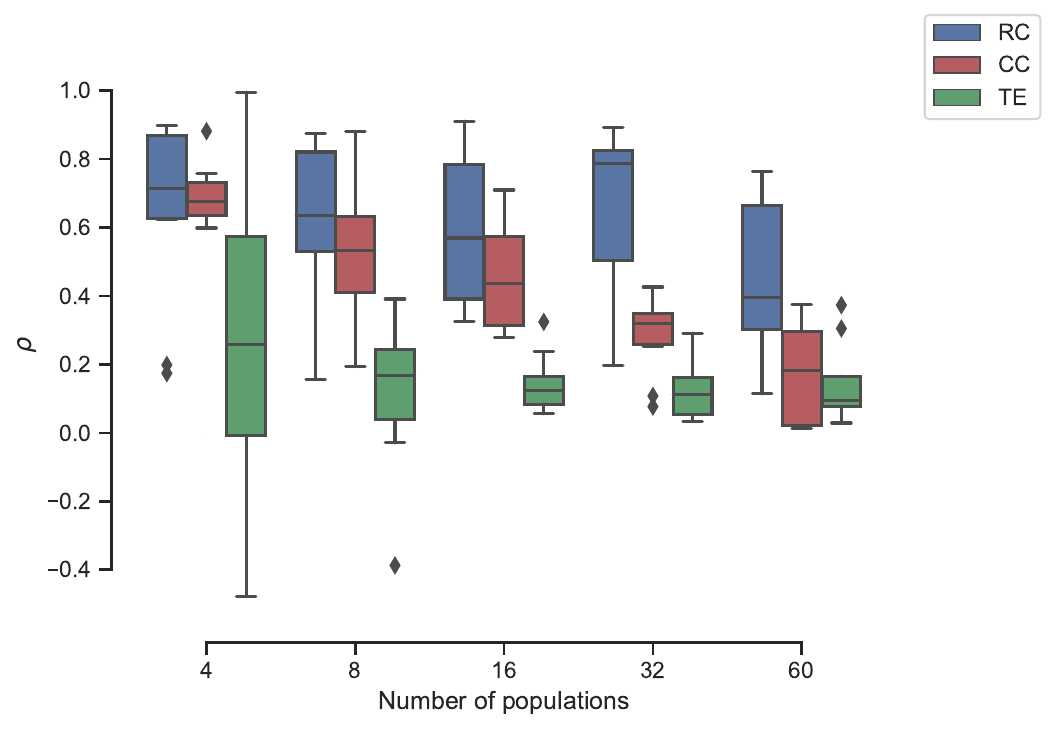}
        \caption{}
        \label{fig:connectivity pearson vs population}
    \end{subfigure}
    \begin{subfigure}[b]{0.48\textwidth}
        \includegraphics[width=\textwidth]{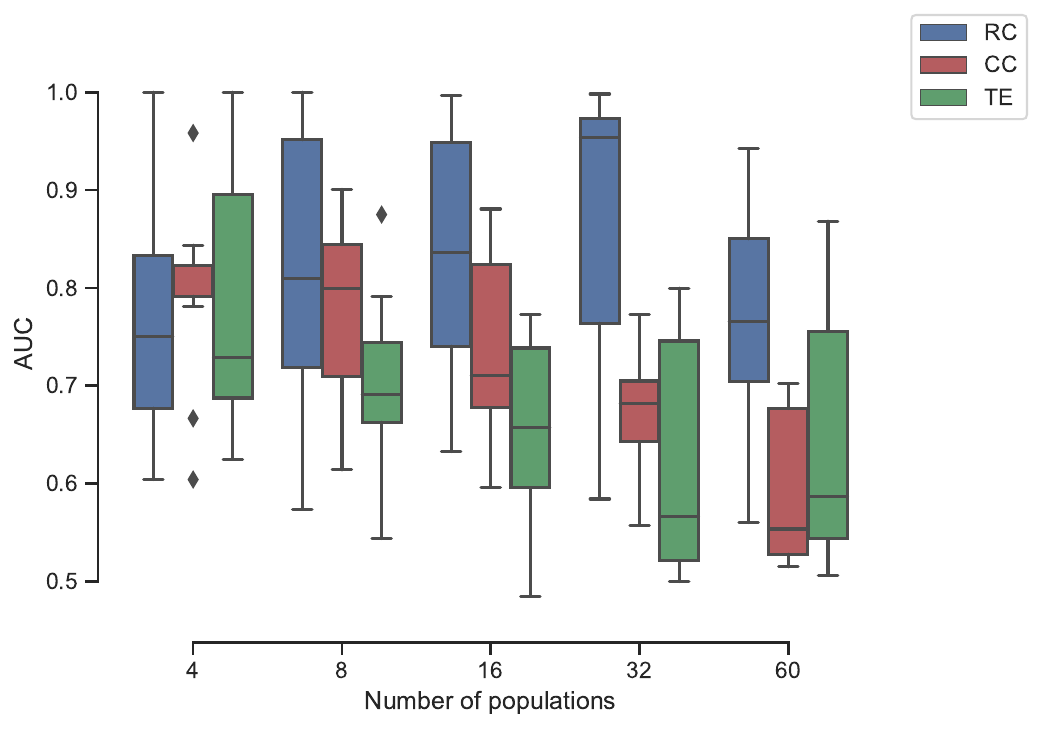}
        \caption{}
        \label{fig:connectivity auc vs population}
    \end{subfigure}
    \hfill
    \caption{Connectivity retrieval performance of the various algorithms with respect to the ground-truth network. (a) Pearson correlation ($\rho$) and ROC AUC metrics as a function of the richness training data set parameter $q$ (Eq. \eqref{eq: q}) for the RC model. Points refer to various network realizations, the lines are a fit to the data (AUC: $R^2 = 0.38$ and $\rho$: $R^2 = 0.47$), the bands refer to the 95 \% confidence interval. (b) Comparison between the performances of the different algorithms evaluated on all the 51 different network realizations. RC refers to the RC model, CC to the Cross-Correlation method and TE to the Transfer Entropy method. Box plots are used to visualize the distribution of the data (minimum value, 1st quartile, median, 3rd quartile, and maximum value). Diamonds are out-layer points. To gauge the statistical relevance of the algorithm comparison, two-sided Mann-Whitney-Wilcoxon test with Bonferroni correction was applied; AUC-RC vs. AUC-CC: $U-stat. = 1820$, $p-value = 4.527\times 10^{-3}$; $\rho$-RC vs. $\rho$-CC: $U-stat. = 1864$, $p-value = 1.568\times 10^{-3}$; AUC-RC vs. AUC-TE: $U-stat. = 1993$, $p-value = 4.402\times 10^{-5}$; $\rho$-RC vs. $\rho$-TE: $U-stat. = 2417$, $p-value = 2.534\times 10^{-12}$; (c),(d) Comparison between the performances of the different algorithms as a function of the number of populations (network nodes), where each population represents a circuit of 5 neurons.}
    \label{fig: connectivity results}
\end{figure}

Furthermore, we compared the RCN model’s performance with the Cross-Correlogram (CC) \cite{perkel1967neuronal,salinas2001correlated} and Transfer Entropy (TE) \cite{PhysRevLett.85.461} methods. These methods are often used to analyze neuronal interactions and connectivity patterns. Our investigation spanned various network configurations, (a total of 51 simulations), including different network sizes, morphologies (sparse vs. dense populations), and ratios of inhibitory to excitatory connections. These structural variations led to significant differences in electrophysiological dynamics, such as spike and burst rates. To gauge the RC model’s effectiveness, we introduced a parameter $q$, which reflects the richness of the training data set:
\begin{equation}
    q \triangleq \frac{N_{t}}{N_{ch}}
    \label{eq: q}
\end{equation}
where, $N_t$ is the total number of time steps used in the training of the RC model and $N_{ch}$ is the number of population clusters (nodes) in the ground-truth network or the number of MEA channels in the experimental data. Fig. \ref{fig: connectivity results} summarizes the connectivity retrieval and validation outcomes. In Fig. \ref{fig:connectivity auc & rho vs q}, we observed a notable correlation between the connectivity retrieval performance and the parameter $q$. This suggests that for larger networks, more time steps in the dynamics need to be included during training. Additionally, the predictive ability of the RC model is influenced by the sparseness of neuronal populations. Figure \ref{fig:connectivity general, algorithms} clearly demonstrates that the RC model consistently outperforms the CC and TE methods in terms of connectivity retrieval. Further insights are provided by Figs. \ref{fig:connectivity pearson vs population} and \ref{fig:connectivity auc vs population}. These figures break down algorithm performance with respect to network size, considering $\rho$ and $AUC$, respectively. Notably, as the number of populations increases, the competing algorithms (CC and TE) exhibit a decline in performance. In contrast, the RC model remains less sensitive to network size, indicating its robustness across varying scales.

\subsection{Prediction of spatio-temporal response of the network}
\label{sec: Results- stim prediction}
Here, we assess the performance of the RC model in estimating the response of a neuronal network when presented with specific inputs (stimulation). The RC model was trained based on the neuronal spontaneous activity. After training, we posses the operator $\hat{\mathcal{F}}$ (Eq. \eqref{eq: full}), which characterizes the network dynamics. Our hypothesis is that the trained RC model can also predict the network's response to localized stimuli. To facilitate the discussion, we label the network state as $\Tilde{\mathbf{u}}[n] \equiv \Tilde{\mathbf{y}}[n]$. Given an initial state $\Tilde{\mathbf{u}}[0] = \mathbf{u}_{in}$, representing a locally stimulated neuronal population, we compute the predicted response of the network $\Tilde{\mathbf{U}}\in \mathbb{R}^{N_{ch}\times N_t}$, by propagating the initial state for a number of time steps using the trained RC model operator $\hat{\mathcal{F}}$. To evaluate the prediction, we estimate the error between the predicted response $\Tilde{\mathbf{U}}$ and the observed response $\mathbf{U}$, obtained either from the ground-truth network or through experiments. For this evaluation, we prepared a specific test data $\mathbf{U}$ for each tested neuronal network (both experimental and simulated), corresponding to a localized stimulus applied in the vicinity of a specific neuronal population (in the experimental context- around a specific electrode).

The evaluation of the prediction accuracy is performed by a ROC curve analysis to spatially discriminate between responsive and unresponsive neuronal populations (nodes). We also evaluate prediction accuracy based on temporal state evolutions. Specifically, we define the error $\Bar{R}$ as:
    \begin{equation}
    \label{eq: R bar}
        \Bar{R} = \sum_i \chi_i \min_\tau \left( \sqrt{\sum_j \upsilon_{i,j,\tau} \bigl(U_{i,j} - \Tilde{U}_{i,j-\tau}\bigr)^2} \right) \;,
    \end{equation}

where $\chi_{i}$ and $\upsilon_{i,j,\tau}$ are the spatial and temporal weighting coefficients, respectively, (defined in Eqs. \ref{eq: chi} and \ref{eq: temporal weights}, respectively); $\tau \in [\tau_{min},\tau_{max}]$ is the time lag between the predicted and the observed responses. $\Bar{R}$, described in details in the Section \ref{sec: response test}, quantifies the accuracy of the trained RC model in capturing both the amplitude and time lag aspects of the actual responses. \ref{fig:Stim1} illustrates the concept of only spatial and spatio-temporal predictions.

In Section \ref{sec: connectivity results}, we benchmarked the model’s performance using synthetic data. In this section, we expand our evaluation to include not only synthetic data but also experimental data recorded from cultured mice cortical neurons via MEA. The NEST simulation data plays a crucial role in enhancing our datasets and enabling a comprehensive assessment of response prediction.
The outcomes of the response test, derived from the utilization of NEST simulation data, are presented in Figs. \ref{fig: response test example simulation} and \ref{fig: Simulations response test}. The former showcases specific results for a particular network, while the latter provides a summary of performance metrics. Similarly, Figs. \ref{fig: response test example experiment} and \ref{fig: Experimental response test} exhibit similar results obtained from MEA experimental data. Table \ref{tab: Respose prediction metrics table} summarizes the evaluation of performance metrics.

\begin{table}[h]
\centering
\renewcommand{\arraystretch}{1.5}
\caption{Performance metrics for the response prediction test. $\Bar{R}$ represents the root-mean-squared-errors, assessing the accuracy of spatio-temporal prediction while AUC denotes the area-under-the-curve in ROC analysis, evaluating the accuracy of the spatial prediction. For each experiment, we employed either a simulated network or a biological culture, subjecting them to various stimulation protocols. The datasets originate from both NEST simulations and MEA experiments.}
    \begin{tabular}{||c|c||c|c||}
    \hline
    \multirow{2}{*}{\textbf{Data type}} & \multirow{2}{8em}{\centering \textbf{\# Protocols (out of \# experiments) }} & \multicolumn{2}{c||}{\textbf{Metric} [mean $\pm$ std]}\\
        \cline{3-4}
         & & $\overline{R}$ & $AUC$\\
        \hline
        \hline
         NEST Simulation & 183 (51) & $0.195 \pm 0.141$ &  $0.876 \pm 0.222$\\
          MEA Experiment & 27 (15) & $0.104 \pm 0.054$ & $0.754 \pm 0.160$\\ [1ex]
          \hline
    \end{tabular}
    \label{tab: Respose prediction metrics table}
\end{table}

\subsubsection{Results for the NEST simulation training data}
\label{sec: simulations response test}
\begin{figure}[h!]
    \centering
    \begin{subfigure}[b]{0.65\textwidth}
        \includegraphics[width=\textwidth]{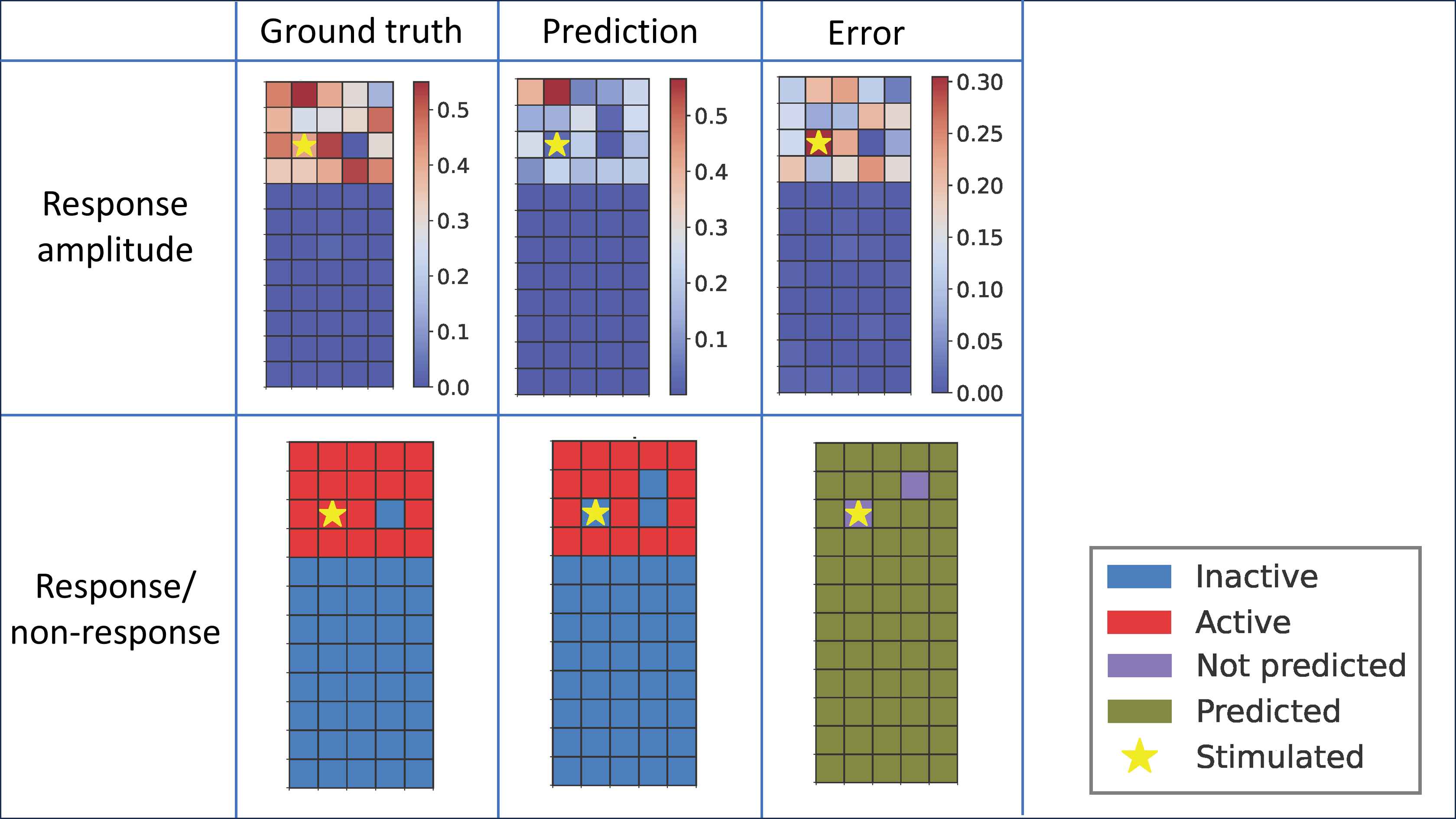}
        \caption{}
        \label{fig:response maps sim}
    \end{subfigure}
        \begin{subfigure}[b]{0.3\textwidth}
        \includegraphics[width=\textwidth]{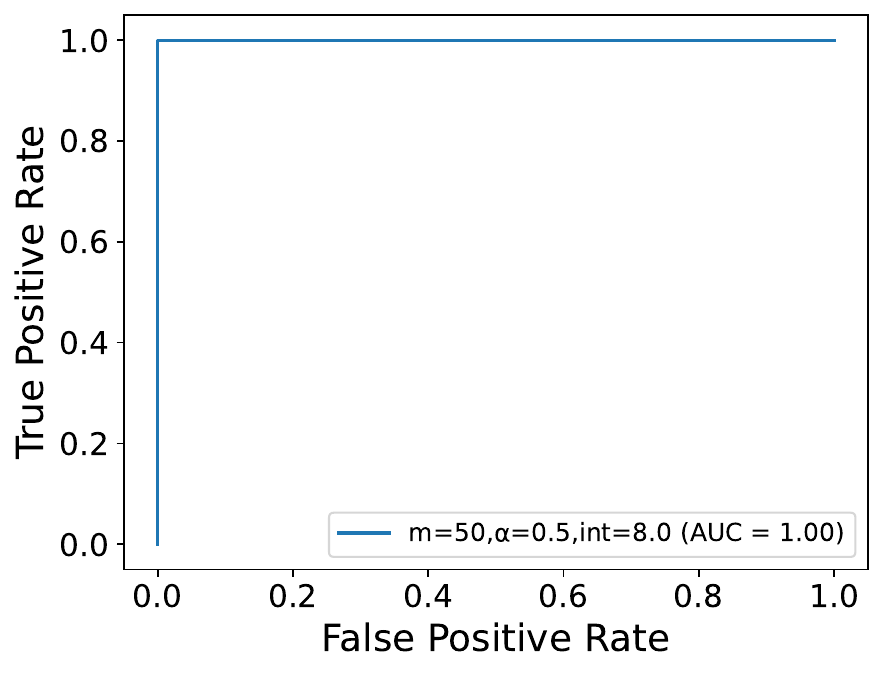}
        \caption{}
        \label{fig:response ROC sim}
    \end{subfigure}
    \caption{Results of a response test for a specific stimulation protocol on the same network as Fig. \ref{fig: CM example}. The RC model was trained on NEST simulation data with a memory strength $\alpha = 0.5$ and a reservoir dimension $m=50$. A stimulus was given on the channel (node) labelled by a yellow star in the maps. (a, top panels) The amplitude response map for the ground truth network (left), the amplitude response map predicted by the RC model (centre) and the error maps for the RC model predictions as given by $\varepsilon$ (Eq. \eqref{eq: channel error} (right). The color maps refer to the amplitudes of the channel (node) response or to the error value. (a, bottom panels) Ground truth network's channel activity map (left), RC model node activity prediction map (centre), and error map evaluated as activity prediction correctness (right). The color code is given in the inset. (b) The ROC curve.}
    \label{fig: response test example simulation}
\end{figure}

The comparison of the responses of the networks in Fig. \ref{fig: CM example} to a local stimulus is shown in Fig. \ref{fig: response test example simulation}. The stimulation protocol excites the neuronal population probed by the 11th channel (as referenced in Fig. \ref{fig: CM example}). The ground-truth data for the test set was extracted from the NEST simulation, while the RC model-predicted data was generated by operating the trained RC model with an input vector $\mathbf{u}_{in}$ and propagating it for 20 time steps. Fig. \ref{fig:response maps sim} illustrates that only one of the three uncoupled groups yields a response for both models. Specifically, the spatial distribution of the response amplitudes (top panels in Fig. \ref{fig:response maps sim}) shows a proper distribution with varying amplitudes and temporal profiles (see Fig. \ref{fig:channel predictions sim}). These differences are quantified in the error maps, where each node displays the results of the error estimations (Eq. \eqref{eq: channel error}). A simpler comparison can be performed by a binary assessment of predictions, which channels are active or resting (bottom panels in Fig. \ref{fig:response maps sim}). The corresponding ROC curve analysis, depicted in Fig. \ref{fig:response ROC sim}, yields $AUC=1$. More details on the analysis are provided in Section \ref{sec: Methods}.

\begin{figure}[h!]
\centering
    \begin{subfigure}[b]{0.35\textwidth}
        \includegraphics[width=\textwidth]{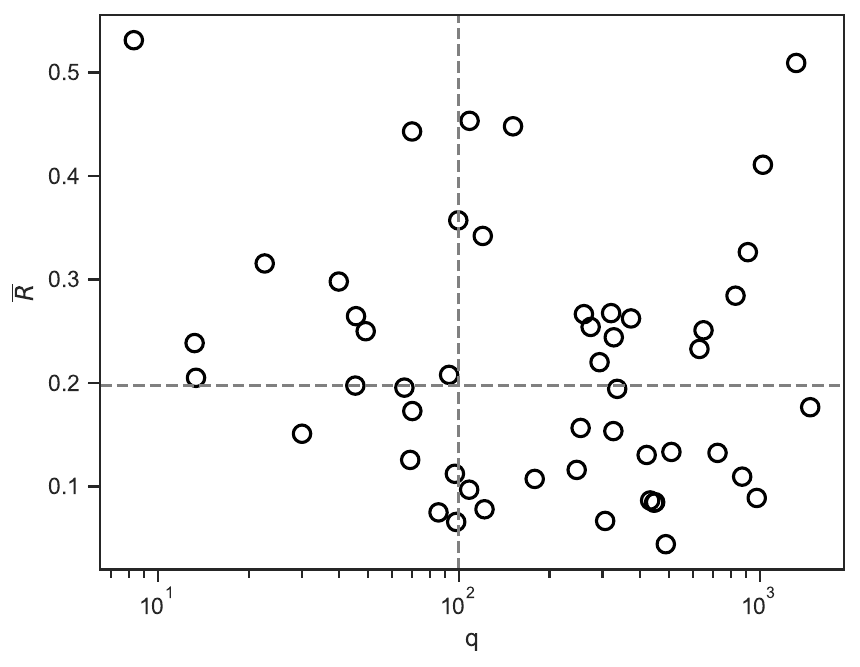}
        \caption{}
        \label{fig: response test simulation R_bar vs q}
    \end{subfigure}
    \begin{subfigure}[b]{0.35\textwidth}
        \includegraphics[width=\textwidth]{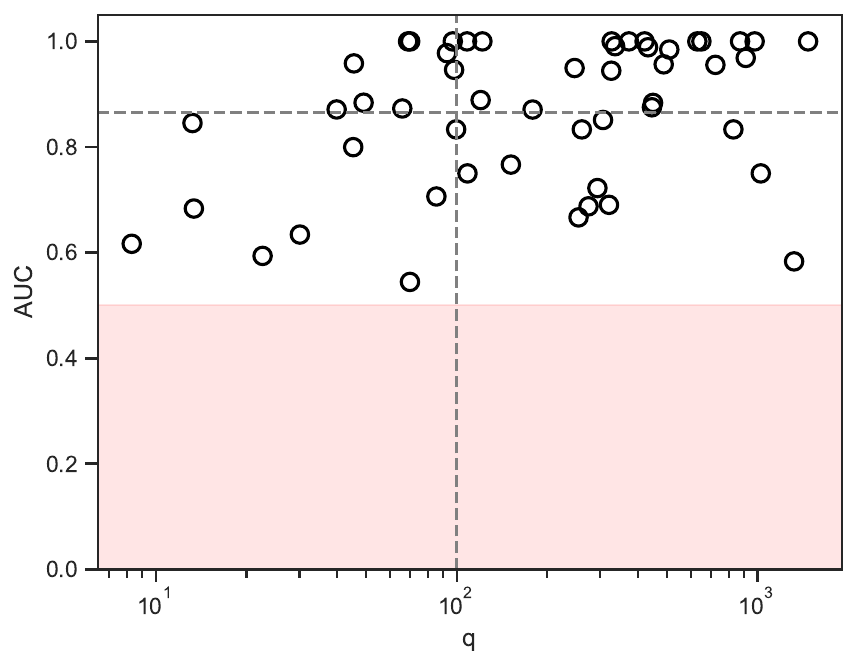}
        \caption{}
        \label{fig:response test simulation auc vs q}
    \end{subfigure}
    \begin{subfigure}[b]{0.35\textwidth}
        \includegraphics[width=\textwidth]{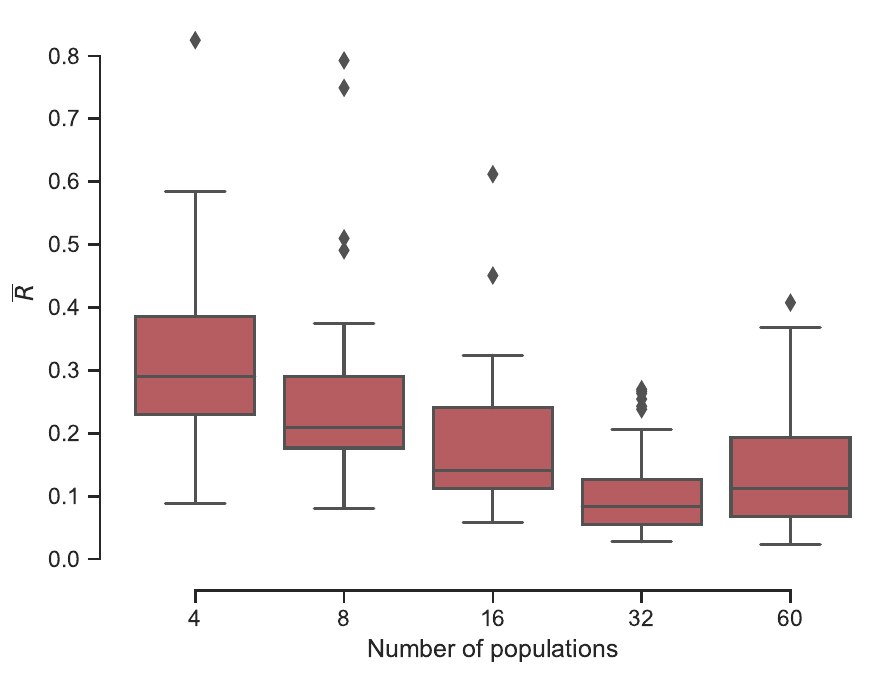}
        \caption{}
        \label{fig: response test simulation R_bar vs population}
    \end{subfigure}
    \begin{subfigure}[b]{0.35\textwidth}
        \includegraphics[width=\textwidth]{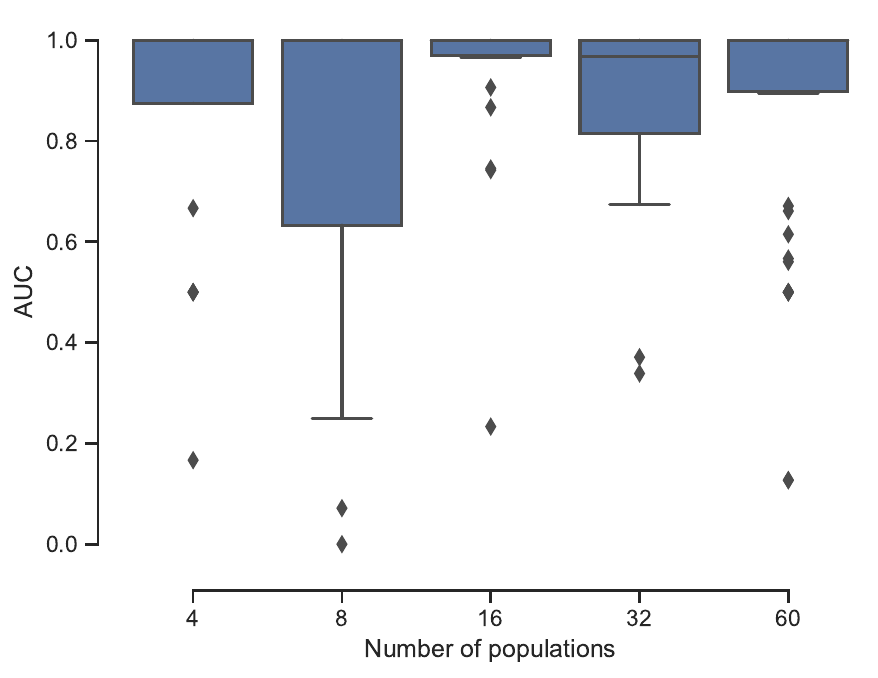}
        \caption{}
        \label{fig: response test simulation auc vs population}
    \end{subfigure}
    \begin{subfigure}[b]{0.35\textwidth}
        \includegraphics[width=\textwidth]{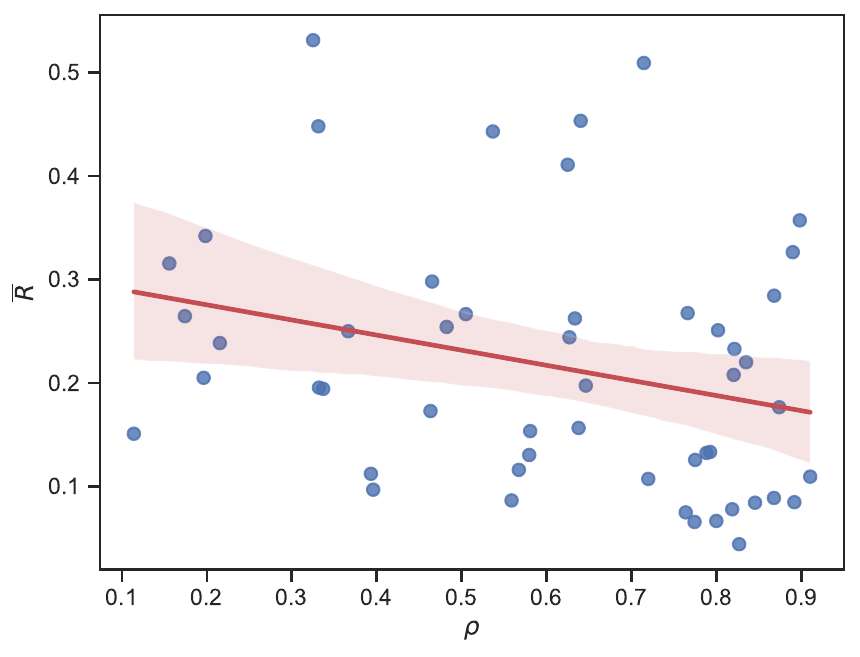}
        \caption{}
        \label{fig: response test simulation R_bar vs rho_conn}
    \end{subfigure}
    \begin{subfigure}[b]{0.35\textwidth}
        \includegraphics[width=\textwidth]{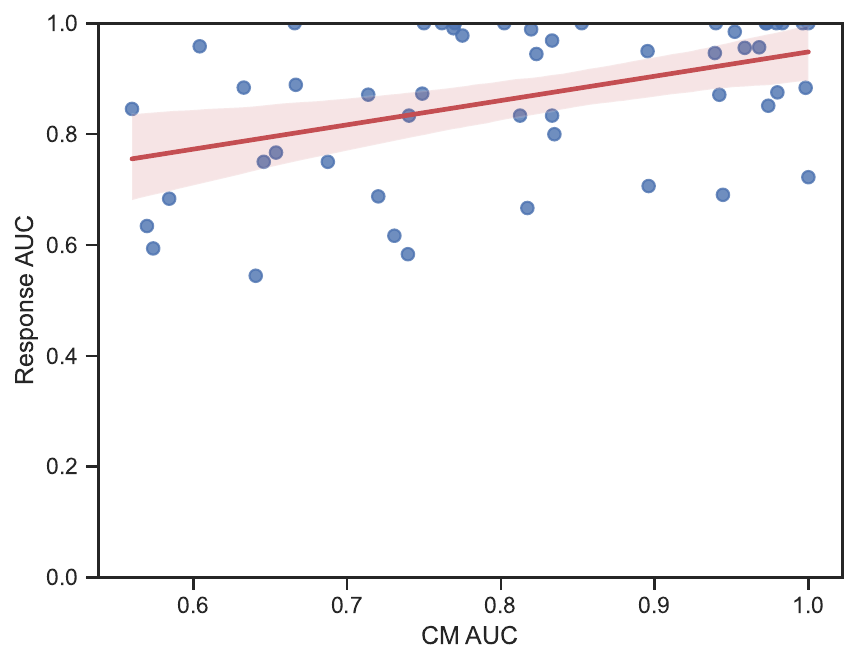}
        \caption{}
        \label{fig: response test simulation auc_resp vs auc_conn}
    \end{subfigure}
        \caption{Performance of the RC model for the network response prediction test (using NEST simulation training data). (a), (c) and (e) present the $\Bar{R}$ metric (spatio-temporal response)- lower values are indicative of a more accurate prediction; (b), (d) and (f)- AUC metric (spatial response)- values closer to 1 indicate more accurate prediction. (a), (b) the metrics as a function of $q$ (Eq. \eqref{eq: q}). The horizontal gray dashed line indicates the mean value of the associated metric, the vertical line is at $q=100$, which we set as an arbitrary boundary between dense and sparse network. Each dot corresponds to the average over different protocols of a specific network. The red zone in (b) indicates the range where the predictions fall below the level of random predictions ($AUC<0.5$). (c), (d) box plots of the metrics for different simulated network size. (e), (f) the response test results against the connectivity retrieval results of the same network. The red lines refer to a least square fit ($\Bar{R}$ vs. $\rho$: $R^2 = 0.060$; Response AUC vs CM AUC: $R^2 = 0.176$), the bands refer to the 95 \% confidence interval.}
        \label{fig: Simulations response test}
\end{figure}

Similar response prediction tests were conducted on all the networks discussed in Section~\ref{sec: connectivity results} by employing various stimulation protocols for each network configuration. Figure~\ref{fig: Simulations response test} provides an overview of the results. Figures~\ref{fig: response test simulation R_bar vs q} and~\ref{fig:response test simulation auc vs q} show the results as a function of the parameter $q$ (Eq.~\eqref{eq: q}). Each data point represents the average value across different stimulation protocols applied to a specific network. The horizontal dashed lines illustrate the overall average, as presented in Table~\ref{tab: Respose prediction metrics table}. Additionally, a vertical dashed line is arbitrarily placed at $q=100$ to distinguish between dense and sparse networks. The red zone in Figure~\ref{fig:response test simulation auc vs q} represents the region between a completely incorrect prediction ($AUC=0$) and a random prediction ($AUC=0.5$). Notably, the RC model consistently outperforms this “red zone.” Furthermore, there is a discernible trend suggesting that better performance is more likely for higher values of $q$. Considering the results as a function of the network size (Figures~\ref{fig: response test simulation R_bar vs population} and~\ref{fig: response test simulation auc vs population}), this suggests that better results are obtained with longer training periods. Finally, we examined the influence of the accuracy in the connectivity map retrieval on the response prediction test (Figures~\ref{fig: response test simulation R_bar vs rho_conn} and~\ref{fig: response test simulation auc_resp vs auc_conn}). A correlation between a low $\Bar{R}$ value for a large Pearson correlation coefficient ($\rho$) or between a high $AUC$ value of the response test and a high $AUC$ of the connectivity retrieval is clear.

\subsubsection{Results for the MEA experiments training data}
\begin{figure}
\centering
    \begin{subfigure}[b]{0.65\textwidth}
        \includegraphics[width=0.98\textwidth]{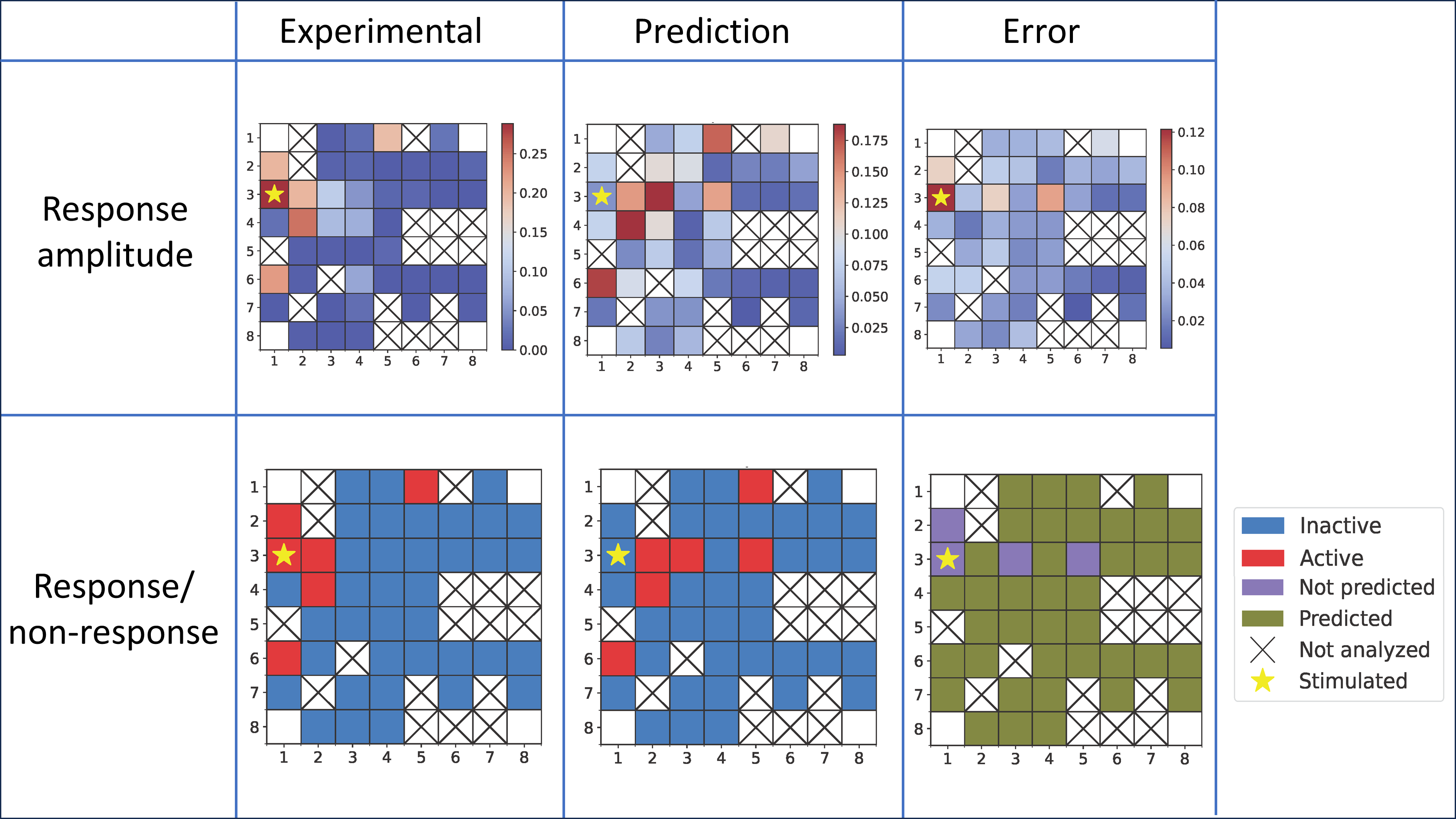}
        \caption{}
        \label{fig:response maps exp}
    \end{subfigure}
        \begin{subfigure}[b]{0.3\textwidth}
        \includegraphics[width=0.95\textwidth]{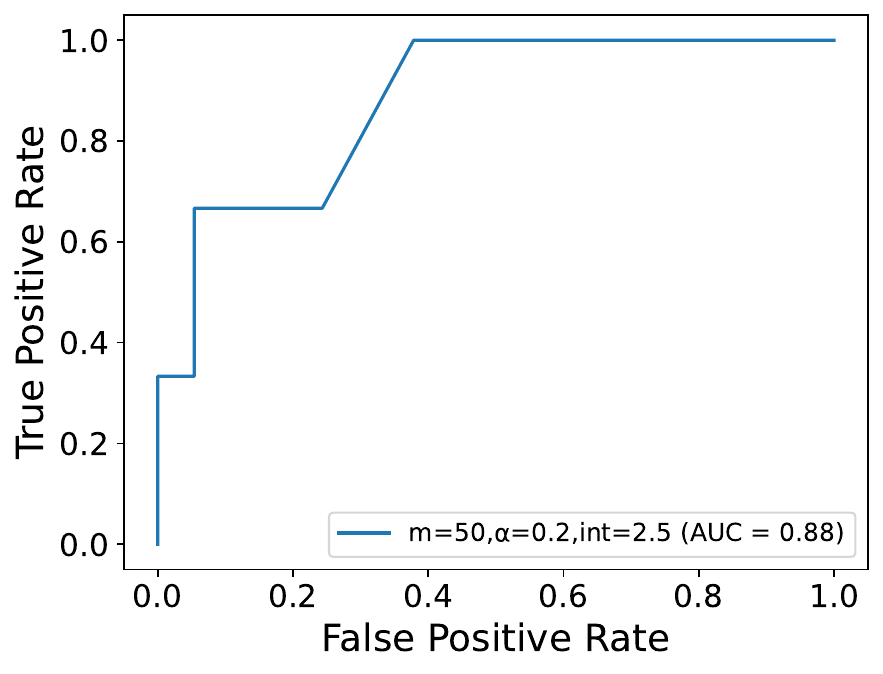}
        \caption{}
        \label{fig:ROC exp}
    \end{subfigure}
    \begin{subfigure}[b]{\textwidth}
    \centering
        \includegraphics[width=0.9\textwidth]{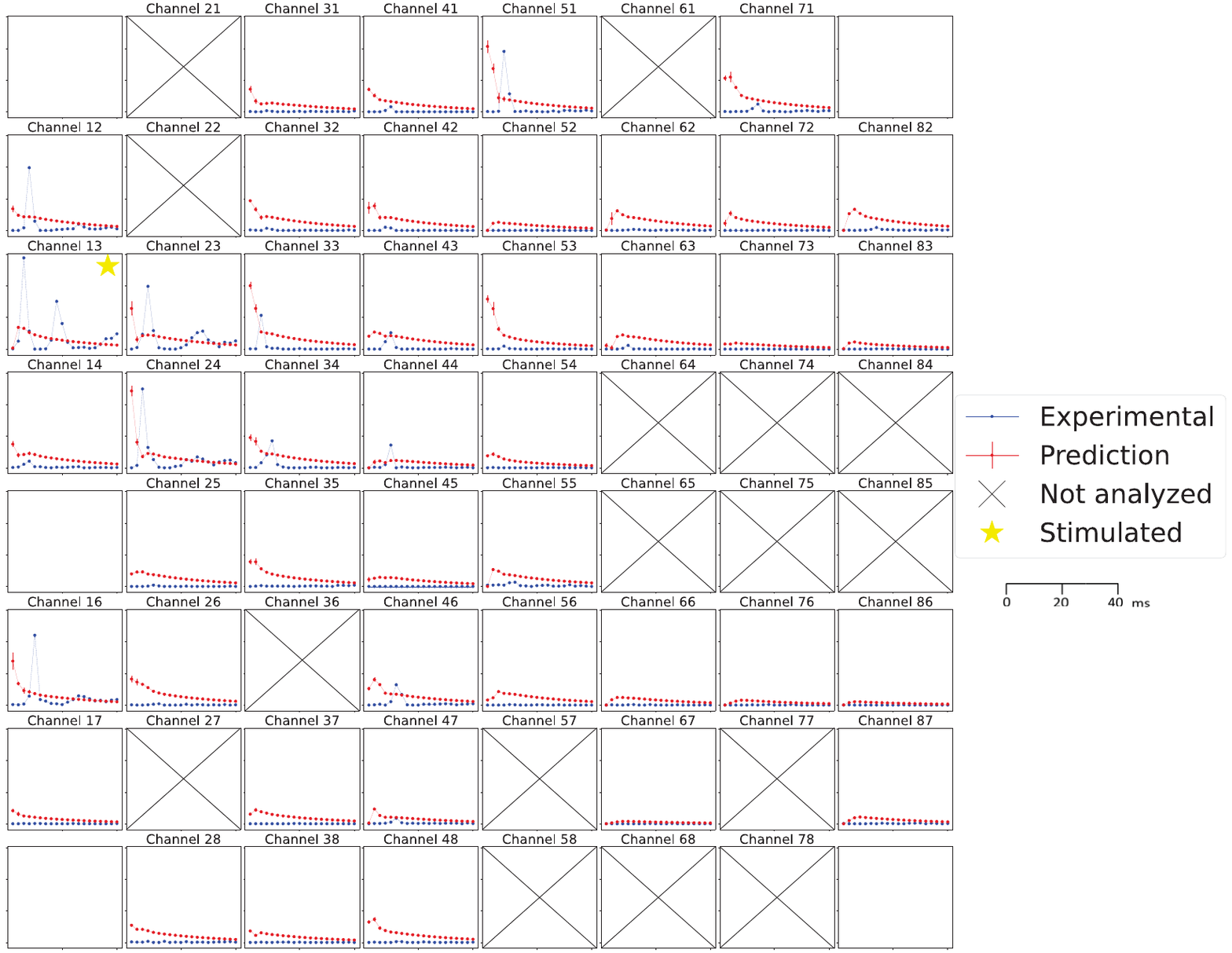}
        \caption{}
        \label{fig:channel predictions exp}
    \end{subfigure}
    \caption{Response test to a specific stimulation protocol for the neuronal culture analyzed in Fig. \ref{fig: CM example}. The RC model
was trained on a 60 electrodes MEA recording data with a memory strength $\alpha = 0.2$ and a reservoir dimension $m = 50$. (a) Activity maps for the experiment (left), for the predicted response (center) and for the error (right). The same description as in Fig. \ref{fig:response maps sim} applies. (b) The ROC curve. (c) Each node (channel for experiment) temporal response to a stimulus given at the node (channel) 14 (labeled by a yellow star). The time scale ($x$ axis) is given in the legend. Data are sampled with a time step of 2 ms. The $y$ axis is the normalized ISR in arbitrary units, scaled identically. From these time profiles $\Bar{R} = 0.047$ and $\Bar{\tau}_l = 6.46ms$ were estimated.}
    \label{fig: response test example experiment}
\end{figure}

\begin{figure}
\centering
    \begin{subfigure}[b]{0.48\textwidth}
        \includegraphics[width=\textwidth]{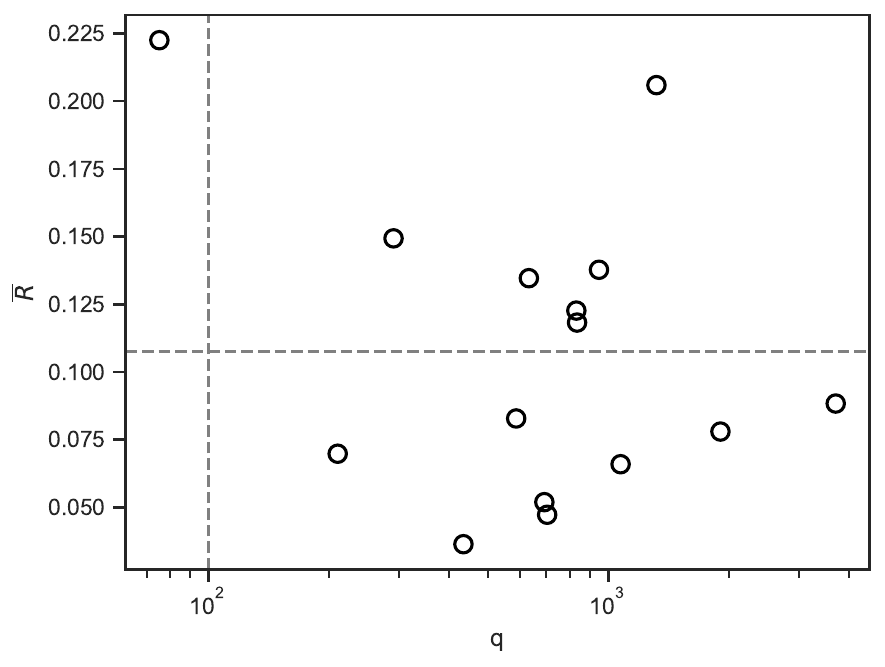}
        \caption{}
        \label{fig: response test experimental R_bar vs q}
    \end{subfigure}
    \begin{subfigure}[b]{0.48\textwidth}
        \includegraphics[width=\textwidth]{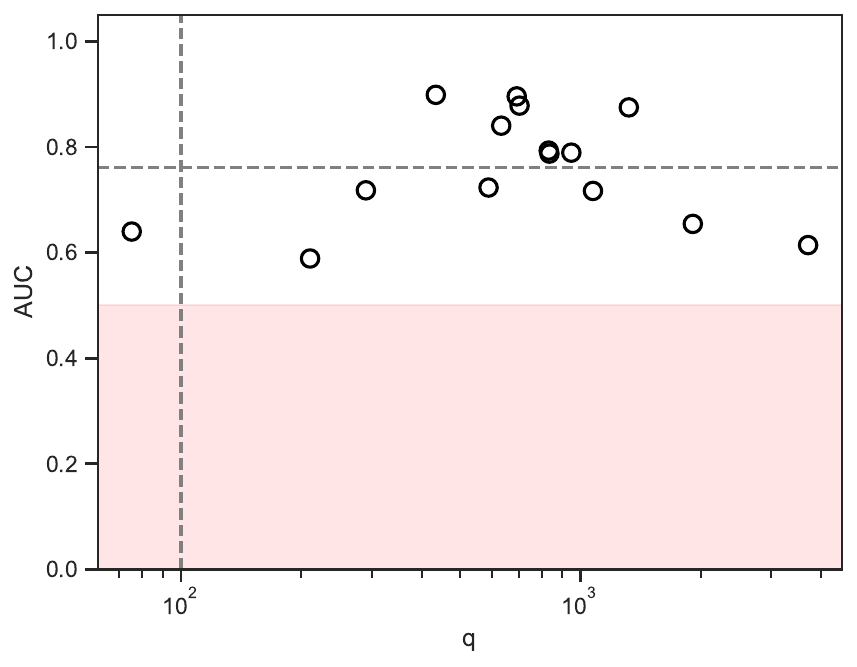}
        \caption{}
        \label{fig:response test experimental auc vs q}
    \end{subfigure}
    \caption{Performance of the RC model for the response prediction test applied to neuronal cultures (MEA experiments data). (a) and (b) the $\Bar{R}$ and $AUC$ metrics as a function of $q$ (Eq. \eqref{eq: q}). The horizontal gray dashed line indicates the mean value of the associated metric, the vertical line is at $q=100$, which we set as an arbitrary boundary between dense and sparse neuronal cultures. The red zone in (b) indicates the range where the predictions fall below the level of random predictions ($AUC<0.5$).
    \label{fig: Experimental response test}}
\end{figure}

A neuronal culture was measured using a 60-electrode MEA. Utilizing the basal activity, an ICM was retrieved as shown in Fig.~\ref{fig:CM exp}. Subsequently, a stimulation test was conducted by exciting the neuronal culture using a specific electrode or a localized spot illumination (see Supplementary Materials), and recording the resulting map with the MEA. Simultaneously, a predictive response test was performed using the trained RC model, where the same network node was excited as discussed in Section~\ref{sec: simulations response test}. A comparison between the experimental MEA data and the RC model predictions is presented in Figs.~\ref{fig: response test example experiment} and~\ref{fig: Experimental response test}. Notably, the RC model accurately predicts the spatio-temporal response of the neuronal culture to a localized stimulus ($\Bar{R} = 0.047$). This is evident in the activity maps and the single-node temporal evolutions (single-channel recordings for the experiment) shown in Fig.~\ref{fig: response test example experiment}. The ROC analysis yields a high $AUC$ value of 0.88 in this specific case. Furthermore, repeated experiments with different neuronal culture preparations and stimulation protocols were analyzed using the RC model, and prediction tests were performed. Overall, the RC model captures the essence of the neuronal culture interconnections, accurately predicting how localized excitation induces a network response. Quantitative assessment of the predictions using the $\Bar{R}$ and $AUC$ performance metrics is shown in Fig.~\ref{fig: Experimental response test} and summarized in Table~\ref{tab: Respose prediction metrics table}. It is observed that the RC model accuracy, as a function of the $q$ parameter, yields better results for an average network size and testing time. Overall, lower $\Bar{R}$ and $AUC$ values with respect to the case of synthetic data are observed which indicates a proper description of the active channel temporal signals (as indicated by the $\Bar{R}$ value) and a partial capture of the intricate network in the neuronal culture (as indicated by the $AUC$ value).

\section{Discussion}
In this study, we developed a computational model that decodes spatio-temporal data from electrophysiological measurements of neuronal cultures. The model reconstructs the network structure on a macroscopic domain and predicts the response to a localized stimulus. Our primary goal was to create an advanced experimental data analysis tool for processing complex time-series. The results obtained indicate that the model not only serves as a data analyzer but can also function as a network simulator. In the following sections, we delve into the model's fundamentals, provide insights into the computational processes, and openly address some of the limitations encountered during our research.

The model’s core concept is rooted in the neural rate-coding of information paradigm, as proposed in \cite{zador1997spikes}. According to this paradigm, the firing rate of action potentials serves as the primary carrier of information in communication between neuronal cells. The second fundamental assumption underlying the model pertains to the validity of electrophysiological measurements of large neuronal populations, particularly extracellular measurements, to capture their network structure. These measurements have been instrumental in validating the model.

Regarding its computational structure, the model is built upon a recurrent neural network (RNN) with a reservoir computing network (RCN) architecture. The rationale for choosing this type of neural network is straightforward. Numerous studies have demonstrated that biological neuronal networks (such as the brain cortex) exhibit complex computational capabilities. Specifically, temporal data is integrated in a recurrent manner, leading to state-dependent synaptic transmission \cite{mastrogiuseppe2018linking,mante2013context,laje2013robust,enel2016reservoir,cai2023brain}. In other words, the network memorizes past impulses and dynamically adjusts its synapses accordingly. The timescale of this process varies based on the network’s functionality. RCN, inspired by biological systems, incorporates neural recurrence and nonlinearity. An additional advantage lies in its training simplicity: only the output layer requires training. This output layer can be fully linear (although not necessarily), as the nonlinearity is embedded within the reservoir neurons. Utilizing a linear output layer enables cost-effective training through linear regression, as compared to more computationally intensive multilayered recurrent neural networks.

RCNs have been widely used for studying electrophysiological signals from neuronal networks \cite{dockendorf2009liquid,sussillo2009generating,yada2021physical,dranias2013short,ju2015spatiotemporal}. In particular, these studies leverage the reservoir-like properties of neuronal networks to adapt them for input-output tasks.
Our methodology, however, takes a different approach. We simulate the neuronal network using the RCN framework to extract network properties from a given culture. This includes analyzing the connectivity map of the network and understanding its macroscopic functionality.
A similar perspective was explored in \cite{gurel2010functional}, where the RCN approach was applied to learn the dynamics of spike trains in neuronal cultures, specifically in recordings from MEAs. In that work, the focus was on predicting spike trains as point-process propagation between distinct input and output nodes, using log-likelihood optimization. Notably, the authors implemented various types of neural networks, including an adaptive reservoir with a nonlinear output layer, which introduces modeling complexity beyond what we discuss in our paper. 

In our research, the RCN is implemented as a rate-coded spiking neural networks, which we train to adapt to time-domain nonlinear synaptic functions. The training data consists of rate-encoded spike trains. This data structure operates under the assumption that spikes propagating within the network carry information within a characteristic time window—unique to each network—defined as the spike-integration time. Additionally, we extract temporal events from the data that exhibit significant spiking activity over short periods (referred to as network bursts). Consequently, the time-series data undergoes substantial dimensionality reduction. Our unique RCN design dissects the network's information processing at each time step (equivalent to one integration time in the real world). This decomposition occurs into two stages:
\begin{enumerate}
    \item Pre-Synaptic Stage: Executed by the input layer and the input of the reservoir layer.
    \item Post-Synaptic Stage: Executed by the output layer and the output of the reservoir layer (see Figs. \ref{fig: artificial neural network} and Figure \ref{fig: Synaptic description}).
\end{enumerate}
It's important to note that this stage sequence corresponds to one time-step and does not necessarily describe the realistic temporal propagation of signals within this time window. The reservoir layer comprises a set of independent micro-reservoirs, equal in number to the macroscopic network nodes (corresponding to the measurement sites or electrodes). In the pre-synaptic stage, we sample the network state independently for each node using its own micro-reservoir. Each micro-reservoir integrates its current state with the input signal and applies a nonlinear function. In the post-synaptic stage, the integrated and processed state of the reservoir (composed of all micro-reservoir states) is linearly mapped to the next network state, taking into account the synaptic coupling between different micro-circuits.

We demonstrated that by adapting the model to replicate the observed dynamics in our experiments, followed by applying a linear approximation of the transfer function, we can uncover the fundamental (or intrinsic) connections between the populations within the network. These intrinsic connections describe the network’s fundamental connectivity, independent of any internal or external stimuli that might modify them.

Interestingly, as indicated by the equations governing the dynamics, these connections change during the full operation of the model, depending on the network’s activity. The resulting connectivity map from this analysis can be referred to as \textit{Effective Connectivity}, in contrast to \textit{Functional Connectivity} \cite{friston2011functional}. The term \textit{Effective Connectivity} emphasizes that these connections are derived from actual causality relations between populations based on electrical signals, as opposed to the statistical relations seen in \textit{Cross-Correlation} or \textit{Transfer Entropy} analysis, which yield \textit{Functional Connectivity}. We assessed the better performance of our model in predicting the connectivity of the network compared to these last methods.

Furthermore, we tested the model's predictive capabilities in terms of input-output prediction. We trained the model using network basal (spontaneous) activity, assuming that this dataset would reveal interactions among network nodes. During the testing phase, we hypothesized that applying local stimulation to the network would elicit a correlated response from the circuits connected to the stimulated population. To assess this hypothesis, we collected distinct datasets: one capturing spontaneous activity and another capturing evoked activity resulting from a local stimulus. The former was used for training and validation, while the latter served as the test dataset. Subsequently, we simulated the model's response to the specific stimulus and evaluated the accuracy of our predictions. We conducted this procedure using both synthetic and experimental data. Our analysis, based on ROC curve metrics, revealed that the model can accurately predict the network nodes that respond to the stimulus in spatial terms. However, we also assessed the accuracy of temporal predictions for each node (electrode). We observed that the temporal profile of the predicted response may be distorted or time-shifted. This discrepancy could be attributed to the fact that certain physical attributes of the stimulus, such as pulse shape (amplitude and time), are not fully considered by the model. Consequently, the temporal dynamics are limited to the time scales of the discretized time-step (defined as the integration time). Nevertheless, in some cases, the temporal predictions achieved high performance. It's important to note that the limitations in achieving precise temporal predictions may not solely stem from the model's constraints but also from our chosen training approach. For example, the simplifying assumption that predicting evoked activity can be accomplished solely by learning from basal activity might prove overly simplistic. Nonetheless, the model is not confined to the specific training and testing approach involving spontaneous and evoked activities. In practical applications, any type of electrophysiological signal data, following the same data collection principles, can be employed to train the model and potentially improve its prediction abilities. For example, utilizing the evoked activity dataset can reveal the connections within the circuitry responsible for responding to the stimulus.

The model’s performance directly depends on the data it is trained on. An essential feature in the data is its spiking modality, which should exhibit fast spiking events at specific time intervals. These events are often seen as synchronous or quasi-synchronous \cite{chiappalone2006dissociated,chiappalone2007network}. By separating the data into temporal windows during these time events and discarding silent phases, each window serves as a data batch. This quality is evident in the inter-spike interval (ISI) histogram, which typically presents a bimodal or multi-modal shape. Another critical factor affecting model performance is the causality between time-steps, discretized by a short-time window (time-bin), which we infer as the spike integration time of each circuit. The model is trained to identify this causality, and the integration time significantly impacts its effectiveness. This implies that the network has a characteristic time constant defining its information processing time. We propose that this integration time can be approximated using attributes from the data, such as the most significant peak in the inter-burst interval (IBI). This feature also relates to the data’s stochasticity, influencing the model’s ability to learn spatio-temporal patterns. However, characterizing the degree of data stochasticity is beyond the scope of this paper.

Another limitation of the model is that biological systems can undergo structural modifications relatively quickly. The data analyzed by the model describe the state of the tested culture within a limited time frame. Consequently, there is a possibility that some information about the tested culture might be overlooked. To address this, when using the model for systems that exhibit rapid variations over time, one can apply short training sessions and observe the system’s development.

Finally, our model was applied to \textit{in-vitro} cultures of neurons, providing a simplified representation of the structure and functionality of these networks in living organisms \cite{gross1977new,chiappalone2019vitro}. This approach aids in breaking down the intricate architecture of the living brain into smaller functional units. We believe that the methodology developed in this study can also be extended to data with higher spatial resolution, such as that obtained from HD-MEA \cite{muller2015high}, allowing for analysis of interactions even at the level of individual neurons. Furthermore, the model is not limited to analyzing only neuronal signals; it can also be applied to various other types of time traces.

\section{Methods}
\label{sec: Methods}
\subsection{The paradigm}
We consider a multi-site measurement of electrophysiological signals from a neuronal culture, such as 2D microelectrode array (MEA). We seek to represent the tested culture as a network where each node corresponds to one measurement electrode. Each electrode samples the electrophysiological signals from the neuron ensemble (consisting of a few neurons) found in its vicinity (Fig. \ref{fig: electrode & neurons}). Therefore, each node has to represent a complex neuronal circuit whose dynamics by itself is driven by numerous interacting neurons. We hence define the domain of the MEA measurement as the \textit{macroscopic domain}, which is the one described by the network. The neuronal structure probed by each electrode will be referred as the \textit{microscopic domain} (or later as the \textit{reservoir domain}) and corresponds to the single node in the network. The \textit{data unit} which is contained in each of these nodes is a sample of the electrophysiological signals measured by one electrode expressed in instantaneous spike-rate measured in a specified time window. The time window is determined by a characteristic information flow rate in the culture. This unit of time is dependent on many properties of the culture such as neuron density in the culture, age of the culture and other \cite{chiappalone2006dissociated}, and it characterizes the signal integration time of each electrode.

Let us represent the macro-domain state of the network at each time step $n = 1,2,3...$ with a vector $\mathbf{y}[n]$, where each component of the vector describes the state of a single node, i.e., $\mathbf{y}[n]$ is the signal representation of each electrode at time $n$. Our aim is to seek the time propagation operator (or a synaptic transmission function) $\hat{\mathcal{F}}$ (Eq. \eqref{eq: full}) transforming state $\mathbf{y}[n]$ to $\mathbf{y}[n+1]$. The operator $\hat{\mathcal{F}}$, which is likely to be non-linear due to the nature of neuronal networks, should describe as closely as possible the experimental observation in the electrophysiological measurements, i.e., we aim to fit a model to an observation which can model or predict the spatio-temporal patterns of the neuronal activity in the culture under test.

We then consider that each node in the macro-domain network represents a complex neuronal signal-processing unit. It arises from the fact that typically every measurement site is surrounded by neurons which may be as many as a dozen. The morphology and functionality of each of these micro-circuits embedded in each node of the macro-domain network cannot be easily obtained from the electrophysiological measurements. Also modeling of such neuronal structures is not as easy and has been studied for decades, with numerous models for different scales of dimensions and time \cite{marder2011multiple,gerstner2002spiking}. 

Hence our approach is to represent each measurement node (electrode) as a gate to a particular neuronal circuit (reservoir), where the signals measured at each node are an outcome of a complex operation involving each circuit and the whole network. We therefore propose the artificial neural network (ANN) structure depicted in Fig. \ref{fig: artificial neural network}. This structure represents a simplification of the neuronal dynamics, where each of the neuronal circuits is a black box, whose morphology and functionality are not known but assumed to be reasonably random. 

As seen in Fig. \ref{fig: artificial neural network} and Figure \ref{fig: Synaptic description}, we assume that the signal sampled at each node is an input to and an output from a higher dimensional domain with specific connectivity and functionality. Each reservoir, associated with a node in the macro-domain, represents a micro-neuronal circuit embedded at each of the measurement sites and has inner interconnections that represent the connectivity of the micro-circuits. Each such circuit performs a nonlinear transformation, creating an updated reservoir state, which on one side is stored as a memory to be integrated into the next time steps, and on the other side is used to form the next state of the macro-domain network.

\subsection{Artificial Neural Network Design}
\subsubsection{Domains and Dimensions}
We denote by $N_{ch}$ the number of nodes in the network, where each node is directly associated with an electrode (or a channel) in the experimental measurement. Assuming that the neurons are uniformly distributed in the culture, we appoint a fixed number of connections between each node and the corresponding micro-circuit, such that for each node of the network there is one micro-reservoir (see Fig. \ref{fig: artificial neural network}). Defining $m$ the dimension of the micro-reservoir, it results that the dimension of the reservoir layer $N_{res}$ is:
\begin{equation}
  N_{res} = m \times N_{ch} .
  \label{eq: dimensions}
\end{equation}
 It follows that each $m$ components in the vector space of the reservoir layer correspond to one node in the macro domain. We may associate $m$ with the relative size of each micro-circuit.
\subsubsection{Input Layer}
The input layer refers to the stage between the macro domain and the reservoir layer. Here we assume that the input state to the reservoir is a linear transformation of the data at the corresponding node, such that each component in the macro-domain transforms directly to $m$ inputs of $N_{ch}$ micro-reservoirs in the reservoir layer, and described the inner dynamics of a single micro-circuit. This is done with the linear transformation described by Eq. \eqref{eq: input}, where the state $\mathbf{y}\in \mathbb{R}^{N_{ch} \times 1}$ is transformed to the input reservoir state $\mathbf{x}_{in}\in \mathbb{R}^{N_{res}\times 1}$ through an input matrix $\mathcal{W}_{in}\in \mathbb{R}^{N_{res} \times N_{ch}}$. Since $\mathcal{W}_{in}$ maps each node to a corresponding micro-reservoir, it is represented by the following matrix:
\begin{equation}
\small{
 \mathcal{W}_{in} = 
\begin{bmatrix}
\left (
\mathbf{w}_{in}^{(1)}
\right )
&  0 & 0
& \cdots & 0\\ 
    0 & 
    \left( \mathbf{w}_{in}^{(2)} \right ) & 0
    & \cdots &  0 \\
    0 & 0 & \left( \mathbf{w}_{in}^{(3)} \right ) & \cdots & 0\\ 
    \vdots & \vdots & \vdots & \ddots & \\
    0 & 0 & 0 &  & \left (\mathbf{w}_{in}^{{(N_{ch})}} \right )
\end{bmatrix}
}
\label{eq: W_in}
\end{equation}
where each $\mathbf{w}_{in}^{(i)}\in \mathbb{R}^{m \times 1}$, $i=1,2,...,N_{ch}$ is a vector with random weights taken from a \textit{normal distribution} (peaked at 0), normalized such that $\|\mathbf{w}_{in}^{(i)}\|^2=1$, which can also be expressed as:
\begin{equation}
   \mathcal{W}_{in}^T \mathcal{W}_{in} = \mathcal{I}_{N_{ch}}
   \label{eq: W_in constraint}
\end{equation}
where $\mathcal{W}_{in}^T$ is the transposed input matrix and $\mathcal{I}_{N_{ch}}$ is the unit matrix of order $N_{ch}$.

\subsubsection{Reservoir Layer}
The reservoir layer contains $N_{ch}$ independent micro-circuits (reservoirs) with $m$ nodes each. Each such reservoir models the neuronal circuit probed by each electrode. This layer has two main functionalities: i) nonlinear time-operator, and ii) reservoir state integrator. The dynamics of the reservoir layer is recurrent. At each time step $n$ the reservoir state $\mathbf{x}[n-1]$ is nonlinearly mapped into a new reservoir state $\mathbf{x}[n]$ (Eq. \eqref{eq: reservoir equation non-linear}). This discrete differential relation provides cumulative data at each time step and carries the temporal memory on the activity of the network. Note that at time step $n$, the reservoir state $\mathbf{x}[n-1]$ undergoes an inner transformation through the reservoir matrix $\mathcal{W}_{res}\in \mathbb{R}^{N_{res} \times N_{res}}$, then integrated with the reservoir input state $\mathbf{x}_{in}$, followed by a nonlinear operation $\mathbf{f}_{NL}$. The operation of  $\mathcal{W}_{res}$ is linear and describes the inner data propagation between the nodes of the reservoir layer. The synaptic morphology of each micro-reservoir is random and since they are independent, there are no coupling elements between them. In addition, we assume that $\mathcal{W}_{res}$ preserves the energy of the state, hence it is represented by a block-diagonal and orthogonal matrix as the following:
\begin{equation}
 \mathcal{W}_{res} = 
\begin{bmatrix}
\left (
\mathbf{W}_{res}^{(1)}
\right )
& 0 & \cdots & 0\\
0 & 
\left ( \mathbf{W}_{res}^{(2)} \right )
 & \cdots & 0\\
\vdots & \vdots &  \ddots &  \\
0 & 0 & &
\left ( \mathbf{W}_{res}^{(N_{ch})} \right )
\end{bmatrix}
\label{eq: reservoir matrix}
\end{equation}
where each $\mathbf{W}_{res}^{(i)}\in \mathbb{R}^{m \times m}, i=1,2...N_{ch}$ is a random-orthogonal matrix. Note that each block acts on its corresponding micro-reservoir state. 

Furthermore, we note in Eq. \eqref{eq: reservoir equation non-linear} the memory parameter  $0< \alpha < 1$, which factorizes the orthogonal matrix  $\mathcal{W}_{res}$. This represents the memory strength, such that the effect of the reservoir state $\mathbf{x}[n]$ on the state $\mathbf{x}[n+k]$ will scale as $\alpha^k$. $\alpha = 0$ indicates that the system is memoryless and the current state at time step $n$ depends only on the input.

The nonlinear function $\mathbf{f}_{NL}$ in Eq. \eqref{eq: reservoir equation non-linear} accounts for the impact of the network saturation and plasticity. Specifically, when subjected to a persistent input, the reservoir tends to saturate, while in the absence of input, it retains an excited state due to the influence of previous signals (plasticity). Note that the plasticity persistence effect depends on the parameter $\alpha$. We tried different nonlinear functions, such as tanh and sigmoid, but the performances of the model showed a weak dependence on the choice of the function.  For the results presented in this paper, we considered:
\begin{equation}
\label{eq: our non-linear function}
    \mathbf{f}_{NL}(\mathbf{\xi}) = \Theta(\mathbf{\xi})\tanh(\mathbf{\xi})
\end{equation}
where $\Theta(\mathbf{\xi})$ is the unit step function and is used to limit the emerging values from the reservoir layer to positive values only. 
Furthermore, we introduced in Eq. \eqref{eq: reservoir equation non-linear} the factor $\hat{\mathcal{S}}$ to give each micro-reservoir neuron a different nonlinear response strength. $\hat{\mathcal{S}}$ does not couple different neurons, $\hat{\mathcal{S}} \in \mathbb{R} ^ {N_{res}\times N_{res}}$ and is a diagonal matrix containing random normally-distributed synaptic strengths on its diagonal.

\subsubsection{Output Layer}
The output layer transforms the reservoir state back to the macroscopic domain and is described by Eq. \eqref{eq: output}. Here, we assume a fully connected layer, such that all the $N_{res}$ reservoir nodes are weighted and connected to all the $N_{ch}$ nodes of the macroscopic network. This layer expresses the synaptic connectivity between the different nodes of the network. It is assumed that this transformation is purely linear, taking into consideration that the overall nonlinearity of the model is dominated by the reservoir layer. 
Unlike $\mathcal{W}_{in}$, $\mathcal{W}_{res}$ and $\hat{\mathcal{S}}$, which are matrices with random and constrained elements, $\mathcal{W}_{out}$ has no constraints on the values of its elements, rather its values are learned during the training by using a Lasso regression (Eq. \eqref{eq: lasso regression}).

\subsection{Data}
\label{sec: data structure}
The model operates on the principles of a rate-coded spiking neural network. For model training, we generated tailored datasets that govern episodes of network bursting (NB) activity, representing them as multidimensional sequences of ISR data. The data in this study were generated through either experimental \textit{in-vitro} measurements, involving the recording of neuronal electrical activity using a 60-channel MEA, or through \textit{in-silico} simulations using the NEST simulator. Both datasets underwent a similar preprocessing, where only significant activities were incorporated into the training dataset. This involved the application of burst and NB detection algorithms \cite{pasquale2010self,bakkum2014parameters}. The experimental data required additional preprocessing before burst detection, entailing the transformation of raw electrical signals into spike trains using filtering and a spike detection algorithm \cite{maccione2009novel}. A comprehensive description of the experimental procedures and the algorithms employed are given in the Supplementary Materials.

\subsubsection{Circuit Integration Time and Data Binning}
The training data was structured as a time sequence representing the network state, encoded in spike-rate values. This approach was based on the assumption that there exists a characteristic range of neural spike integration time (found to be within a timescale of a few milliseconds), defining the information processing duration for individual populations within the neuronal culture (Fig. \ref{fig: integration of circuits}). We empirically demonstrated a close correspondence between the integration time (yielding optimal performance of the RC model) and the peak of the inter-burst-interval (IBI) histogram (Fig. \ref{fig: IBI}) within the 2-10 millisecond range. The spike traces are subsequently segmented into time bins of duration equal to the integration time, where each bin contains the ISR (see details in Sec. \ref{sec: integration time} of the Supplementary Materials). Accordingly, we transform the spike trains data to ISR data:
\begin{equation}
    \mathbf{D}[i,t] \rightarrow \mathbf{\Sigma}[i,n] 
\end{equation}
$\mathbf{D}$ and $\mathbf{\Sigma}$ represent 2D matrices, where the first dimension under index $i$ denotes the channel number, and the second dimension represents the time domain. Specifically, $\mathbf{D}$ serves as the complete spike train dataset, containing firing times denoted as $t_f$:
\begin{equation}
    \mathbf{D}[i,t]= \sum_f \delta(t-t_{i,f})
\end{equation}
$\mathbf{\Sigma}$ is the time-binned spikes matrix, where $n$ serves as the bin index, later identified as the unit time step:
\begin{equation}
    \mathbf{\Sigma}[i,n]=\frac{1}{t_{int}}\sum_{t=t_{n-1}}^{t_n} \mathbf{D}[i,t], \;\; t_n = n\cdot t_{int}
\end{equation}
$t_{int}$ is the integration time (Eq. \eqref{eq: discrete time})
\subsubsection{Training and Validation Datasets}
\label{sec: Training}
Following the steps described in the previous subsection, we generate the training and validation datasets. To ensure a consistent domain for the RC model across all datasets, we standardize the ISR data within the $\mathbf{\Sigma}$ matrix, scaling its values to fall within the 0 to 1 range. Subsequently, we construct the dataset by extracting data batches from $\mathbf{\Sigma}$ in segments from the full-length data, employing the following approach:
\begin{equation}
    \overline{\mathbf{\Sigma}}\Bigl[i,n_j:n_j + N_j + \Delta - 1\Bigr] \rightarrow \mathbf{Y}_j\left[i,1:N_j + \Delta\right] 
\end{equation}
$\overline{\mathbf{\Sigma}}$ is the normalized $\mathbf{\Sigma}$ matrix; $n_j$ refers to the time-bin which indicates the beginning of the $j$-th NB. $N_j$ denotes the duration of the $j$-th NB, specified in the number of time bins. The parameter $\Delta$ is used to accommodate extra time steps, guaranteeing the inclusion of the latter part of the NB. Accordingly, the matrix $\mathbf{Y}_j$ represents the training batch number $j$.

Subsequently, we distribute the batches in a shuffled manner, allocating 85\% of them for training and reserving the remaining 15\% for validation, creating matrices $\mathbf{Y}^{tr}_j$, $\mathbf{Y}^{val}_k$ accordingly.

All batches $\mathbf{Y}^{tr}_j$ are then provided to the Lasso regression algorithm (Eq. \eqref{eq: lasso regression}) as merged sequence $\mathbf{Y}^{tr}$ (merging all matrices $\mathbf{Y}^{tr}_j$ along time dimension). The training however is informed about the beginning of a new batch, by resetting the time step to $n=0$ and the reservoir state to $\mathbf{x}[0]=0$ (Eq. \eqref{eq: reservoir equation non-linear}). In this way, we decouple the effects of one NB on another since the time interval between different NBs is considerably longer than the neuronal interaction time. The remaining validation batches $\mathbf{Y}^{val}$ are then used to evaluate the training, by calculating the validation loss:
\begin{subequations}
\label{eq: Validation loss}
    \begin{align}
        \mathcal{L}^{val} = \frac{1}{N_{ch}} \sum_{i=1} ^{N_{ch}} \ell_i
        \label{eq: validation loss total}\\
        \ell_i = \sqrt{\sum_n \omega_i[n] \left|\Tilde{y}_i^{val}[n]-y_i[n] \right|^2}
        \label{eq: validation loss per channel}\\
        \Tilde{y}_i^{val}[n] = \sum_j \Bar{W}_{out} ^{i,j} \cdot x_j[n]+ \Bar{b}_i
        \label{eq: validation output}\\
        \omega_i[n] = \left\lvert \frac{y_i^{val}[n]+\Tilde{y}_i^{val}[n]}{\sum_n (y_i^{val}[n]+\Tilde{y}_i^{val}[n])} \right \rvert 
        \label{eq: weights of validation}
    \end{align}
\end{subequations}

$y_i^{val},\Tilde{y}_i^{val}$ are the observed and modeled validation data for each node $i$; $\Bar{W}_{out},\Bar{b}$ (Eq. \eqref{eq: validation output}) are the learned elements of $\mathcal{W}_{out}$ (i. e. the trained weights) and of the bias vector $\mathbf{b}$ (Eq. \eqref{eq: lasso regression}). To enhance the robustness of our predictions, we use $\omega$ (Eq. \eqref{eq: validation loss per channel}) in our loss estimations. They mitigate the possible presence of a significant portion of zero sequences in the traces $y,\Tilde{y}$. These zero sequences can potentially lead to overly simplistic predictions. These $\omega$ weights, defined in Eq. \eqref{eq: weights of validation}, assign greater importance to non-zero predictions within the trace, thereby addressing the potential bias introduced by the prevalence of zero sequences and enhancing the sensitivity of the model to more informative elements in the data.

\subsection{Linearized Model and Connectivity Analysis}

\label{sec: Linear model}
If the initial state of the reservoir is $\mathbf{x}[0] = 0$ (resting state) and in the absence of any input $\mathbf{y}$, the reservoir state $\mathbf{x}[n]$ does not change with time, see Eq. \eqref{eq: reservoir equation non-linear}. As a consequence, no dynamics in the network nodes $\mathbf{y}[n]$ is observed, see Eq. \eqref{eq: dynamics}. Let us assume (without the loss of generality) that at a certain time-step $n = 1$ we have a small perturbation, $\mathbf{y}[1]$, such that:
\begin{equation}
\mathbf{x}[1]  =  \mathbf{f}_{NL} \left (\hat{\mathcal{S}}{\mathcal{W}_{in} \mathbf{y}[1]}\right ) \approx \hat{\mathcal{S}}{\mathcal{W}_{in} \mathbf{y}[1]}
\end{equation}
This approximation is valid for any nonlinear function that satisfies $f(\xi)\approx \xi$ for $\xi \ll 1$, e.g., Eq. \eqref{eq: our non-linear function}. If such linear regime is maintained in the following $k-1$ steps, from Eq. \eqref{eq: dynamics} follows that the network output at time step $k+1$ is:
\begin{equation}
\mathbf{y}[k+1] \approx  \sum_{n=1}^k \alpha^{n-1} \cdot \left [\mathlarger{
\mathcal{T}}_{n-1} \right] \: \mathbf{y}[k-n+1]
\label{eq: linear approximation}
\end{equation}
where,
\begin{equation}
\mathcal{T}_p = \mathcal{W}_{out}\hat{\mathcal{S}} \left[\mathcal{W}_{res}\hat{\mathcal{S}}\right]^p\mathcal{W}_{in}
\label{eq: transfer matrix}
\end{equation}
is a $N_{ch} \times N_{ch}$ transfer matrix of order $p$. Note that we omitted the constant bias vector $\mathbf{b}$ from Eq. \eqref{eq: output}, since it describes a constant DC offset, and as it results a posteriori, its value is negligibly small.

After training the RC model network, a proper parametrization of the nonlinear model given by Eq. \eqref{eq: dynamics} is found. It follows that the $p$ transfer matrices \eqref{eq: transfer matrix} contain the connection strengths (elements of the matrices) between the nodes for different $p$ steps. These interactions scale with time according to $\alpha ^p$. Next, if we eliminate the recurrence in the reservoir operation by removing the effect of previous steps (no memory in the reservoir, i.e. $\alpha \to 0$ in Eq. \eqref{eq: linear approximation}), we get:
\begin{equation}
    \mathbf{y}[k+1] = \mathcal{T}_0 \mathbf{y}[k] \; .
    \label{eq: zeroth order}
\end{equation}
We define $\mathcal{T}_0$ the \textit{intrinsic connectivity matrix} since it directly gives the weights between the network nodes for two consecutive temporal states, regardless of the memory stored in the reservoir. Indeed, each matrix element $\mathcal{T}^{i,j}_0$ shows the direct connection $j\rightarrow i$, i.e., from node $j$ at time $n$ to node $i$ at time $n+1$.
The higher order $\mathcal{T}_p$ ($p=1,2,\ldots k$) matrices contain the corrections to the connection weights following the reservoir activation still in the linear approximation. The elements of $\mathcal{T}_0$ express both excitatory (positive values) as well as inhibitory connections (negative values).

\subsection{Testing and Performance Metrics}
Model performance was examined by several validation paths: 1. Evaluating training (Eq. \eqref{eq: lasso regression}) and validation loss (Eq. \eqref{eq: Validation loss} while tuning model parameters $\alpha$, $m$ and the integration time (see Supplementary Materials). 2. Benchmarking the connectivity matrix using synthetic data; 3. Estimating the prediction accuracy of the network spatio-temporal response to a localized stimulus using experimental and synthetic data.

\subsubsection{Connectivity Map}
\label{sec: metrics- connectivity}
By obtaining the \textit{Intrinsic Connectivity Matrix}, $\mathcal{T}_0$ from Eq. \eqref{eq: intrinsic conn matrix}, we build a network graph that describes the interactions between the network nodes. We assume that these connections depict the short-term interaction between the nodes, and, hence, the effective connectivity between the different populations (or circuits) in the neuronal culture. To assess the accuracy of the network graph, we utilized two metrics: i) a binary metric, i. e. the presence or absence of connections, assessed by ROC curves, and ii) a weight prediction metric, evaluated through the \textit{Pearson Correlation Coefficient}, $\rho(X,Y)$. In the ROC analysis, we designated '1' for any non-zero weights in the ground-truth network connectivity matrix, denoted as $\mathcal{T}_{GT}$, and '0' for zero values. In contrast, the connections for the RC model were categorized as '1' or '0' using a range of custom thresholds based on the evaluation of True Positive Rate (TPR) versus False Positive Rate (FPR).
As for the \textit{Pearson Correlation Coefficient}, we calculated $\rho(\mathcal{T}_0,\mathcal{T}_{GT})$, where $\rho$, $\mathcal{T}_0$ and $\mathcal{T}_{GT}$ are transformed into 1D vectors of weights for a specific connection. We benchmarked the RC model against connectivity matrices obtained from the Cross-Correlation (CC) and Transfer Entropy (TE) methods. To obtain the connectivity matrices generated by these methods, denoted as $\mathcal{T}_{CC}$ and $\mathcal{T}_{TE}$ respectively, we employed the SpiCoDyn toolbox, as described in \cite{pastore2018s}. Subsequently, we estimated their ROC curve against the ground-truth network connectivity and calculated the correlation coefficients $\rho(\mathcal{T}_{CC},\mathcal{T}_{GT})$ and $\rho(\mathcal{T}_{TE},\mathcal{T}_{GT})$.

\subsubsection{Response Prediction Test}
\label{sec: response test}
As outlined in Section \ref{sec: Results- stim prediction}, a trained model undergoes a response test where the stimulus is represented as an input vector with dimensions $N_{ch} \times 1$. The elements of the vector correspond to the spatial representation of the macroscopic domain (the electrodes layout); the one-column representation infers that the impulse is given during one time-step persisting for one integration time; and the amplitude of the stimulus corresponds to the normalized instantaneous spike-rate, as was coded in the training data. This representation of the impulse describes the effective pre-synaptic impulse given to a target neuronal population that insists on the electrode-probed region.
As the test data, we prepared datasets from experiments of evoked activities driven by a localized stimulus (optical, electrical, or simulated on NEST). The experimental response is a recorded time trace of the neuronal culture activity within a fixed time window following the stimulus in the form of a time histogram, which is known in the literature as \textit{post-stimulus time-histogram} (PSTH).

The following metrics evaluate the accuracy of the prediction of the RC model network's response:
\begin{enumerate}
    \item ROC Curve: used to assess the accuracy of spatial responses by utilizing a binary classification. This distinguishes between responsive channels/electrodes and unresponsive (resting) ones. In establishing the experiment's ROC curve, we designated a channel as '1' (indicating a responsive electrode) if at least one time bin in the PSTH has a value corresponding to one spike per time bin in at least half of the repeated stimuli; and '0' otherwise. The predicted values were determined as the maximum value among the elements of the matrix $\Tilde{\mathbf{Y}}$, representing the peak of each node's predicted PSTH. Subsequently, the ROC curve analysis was conducted as for the connectivity map analysis.
    \item An aggregated error $\Bar{R}$ based on the root-mean-squared-error: used to evaluate the spatio-temporal prediction accuracy by examining the temporal profile of the response for each channel/electrode or nodes.
\end{enumerate}

Due to the differences between the experimental and the modeled stimuli, we expected inaccuracies in the temporal profile of the node response, such as temporal distortion of the signal or time shift. To take into account this, we used the following procedure to compute $\Bar{R}$. Let $\Tilde{\mathbf{U}} = \bigl [\Tilde{\mathbf{u}}[1], \Tilde{\mathbf{u}}[2], \ldots ,\Tilde{\mathbf{u}}[N_t] \bigr]$ and $\mathbf{U} = \bigl [\mathbf{u}[1], \mathbf{u}[2], \ldots ,\mathbf{u}[N_t] \bigr]$ be the predicted and the experimental PSTH, respectively. $\mathbf{U}$ is normalized by the same normalization factor as the basal activity for a given culture (Sec. \ref{sec: Training}).  We then compute the integrated response for each electrode (approximated by a trapezoid integration):
    \begin{equation}
        \varsigma_i(N_t) = \frac{1}{2} \sum_{n=1}^{N_t-1}
        \left (U_{i,n} + U_{i,n+1} \right ), \quad i=1,2,3...N_{ch}
    \end{equation}
where $U_{i,n}$ represents the $i,n$ elements of the matrix $\mathbf{U}$: the first index is the electrode index and the second index is the time step index. Next, we define a weight for each node/channel, as follows:
    \begin{equation}
        \chi_i = \frac{\max(\varsigma_i,\Tilde{\varsigma}_i)}
        {\sum_{i} \max(\varsigma_i,\Tilde{\varsigma}_i)}
        \label{eq: chi}
    \end{equation}
where $(\varsigma_i,\Tilde{\varsigma}_i)$ are the integrated response for the experimental and the model time traces, respectively. Then, we define $R_i$ the cross root-mean-square- error per channel, which can be calculated by:
    \begin{equation}
        R_i[\tau] = \sqrt{\sum_{n=1}^{N_t} \upsilon_{i,n,\tau} \left(U_{i,n} - \Tilde{U}_{i,n-\tau}\right)^2}
        \label{eq: xrmse}
    \end{equation}
    where $\tau\in [-\tau_{max},+\tau_{max}]$ is the time shift between the experimental time trace $U$ and the predicted time trace $\Tilde{U}$, $\tau_{max} = 10$ is arbitrarily used; $\upsilon_{i,n,\tau}$ is a temporal weight coefficient given by:
    \begin{equation}
    \label{eq: temporal weights}
        \upsilon_{i,n,\tau} = \frac{U_{i,n}+\Tilde{U}_{i,n-\tau}}
        {\sum_n \left(U_{i,n}+\Tilde{U}_{i,n-\tau}\right)} \; .
    \end{equation}
Finally, by finding the time lag in which the error is minimal:

\begin{equation}
    \begin{gathered}
        \varepsilon_i = \min (R_i[\tau]) \\
        \tau_{l,i} = {\arg \min} 
        (R_i[\tau]) \; ,
        \label{eq: channel error}
    \end{gathered}
\end{equation}

we can evaluate the aggregated error $\Bar{R}$ :
    \begin{equation}
        \Bar{R} = \sum_i \chi_i \cdot \varepsilon_i \; ,
        \label{eq: R_bar_detailed}
    \end{equation}
which is Eq. \eqref{eq: R bar}, and the aggregated time lag of the prediction:
    \begin{equation}
        \Bar{\tau_l} = \sum_i \chi_i \cdot \tau_{l,i} \; .
        \label{eq: tau_l}
    \end{equation}

In the evaluation of $\Bar{R}$, we used Eq. \eqref{eq: xrmse} to assess the predictive accuracy of signals featuring a peak and to account for temporal misalignment; the weights $\upsilon_{i,n,\tau}$ serve to diminish the significance of zero or low values, as recurrent predictions in subsequent points are considered trivial; the weights $\chi_i$ assign a greater significance to channels exhibiting stronger responses. 

\subsection{\textit{In-silico} Simulation}
\label{sec: NEST simulation}
As the underlying circuital network of a neuronal culture is not accessed by the experimental recordings of the neuronal electrophysiological signals (MEA recording), an \textit{in-silico} model of a neuronal culture has been developed. The basic network unit of such model is defined as \textit{population}, which is composed of a fixed number of point-process neurons described by Izhikevich equations \cite{izhikevich2003simple}, parameterized to follow the dynamic of regular-spiking neurons \cite{izhikevich2004model}. Each neuron inside a \textit{population} is connected to all neighboring neurons (\textit{intra-population connections}), with no self-connections allowed. The different \textit{populations} are then connected randomly to a variable number of the others (\textit{inter-population connections}). Specifically, a connection between two \textit{populations} $i$ and $j$ is obtained by connecting each neuron of \textit{population} $i$ to $n$ randomly sampled neurons of \textit{population} $j$. This network architecture has been chosen to recreate highly interconnected hubs\cite{antonello2022self}, represented by single populations, which should resemble neuronal assemblies surrounding actual MEA electrodes for which action potentials are recorded. Instead, connections among different populations mimic long-range relationships between neurons. For both \textit{intra-} and \textit{inter-population connections}, different types of synapses — static and plastic \cite{gutig2003learning} — have been tested. The weights are assigned using uniform distributions, with distinct excitatory and inhibitory connection ranges. In particular, inhibitory connection weights have a higher absolute value to accommodate their lower prevalence in the network, maintaining an 80/20\% proportion between excitatory and inhibitory connections.
To replicate the spontaneous activity displayed by neuronal cultures, a piecewise constant current with a Gaussian-distributed amplitude has been injected into each neuron to reproduce the fluctuation in the membrane potential (noise component). Moreover, Poisson spike trains addressed either to all neurons or just one neuron inside \textit{populations}, have been utilized to replicate spikes occurring from neurons in the culture not detected by MEA electrodes (background activity). 
The model has been tuned to reproduce the dynamics exhibited by \textit{in-vitro} neuronal networks. In particular, simulations show a mix of spiking and bursting activities as visible by the raster plot depicted in Fig.\ref{fig: raster}, with an average firing rate in line with the experimental recordings. Moreover, the log-inter-spike-interval distribution (log-ISI) has been calculated for each \textit{population} to monitor the network's dynamic. Populations showed different firing dynamics, resulting in heterogenous log-ISI histograms, with the appearance of bimodal histograms (Fig. \ref{fig: ISI one channel}), as observed \textit{in-vitro} neuronal cultures \cite{selinger2007methods,pasquale2010self}, generally indicating bursting events.
The final output of \textit{in-silico} simulation is obtained by collecting spike times of each neuron in a population and by sorting them temporally, thus obtaining a spike train for each \textit{population}, which resembles the spike train acquired after performing spike detection on the raw multi-unit activity signal sampled by a MEA electrode \cite{obien2015revealing}. Additionally, the connectivity matrix of the links between populations can be obtained. In particular, each entry $i,j$ indicates a link between populations $i$ and $j$, and its weight value is a weighted average of all inter-population connection weights by the firing probability of every source neuron in population $i$. The weighted average has been selected to take into account not just structural connections but also the functional relationships across \textit{populations}. Moreover, to test the ability of the RC model to retrieve the connectivity of the \textit{in-silico} model, networks composed of clusters of populations have been created, where \textit{inter-population connections} were allowed only between \textit{populations} belonging to the same cluster. Furthermore, \textit{populations} whose \textit{inter-population connections} were only inhibitory have been used to evaluate the capacity of the RC model to distinguish between excitatory and inhibitory synapses. 

In addition, on the \textit{in-silico} model localized stimulation simulations have been performed. For each \textit{stimulation protocol}, a constant DC input lasting 2 ms was provided to all neurons of a specific \textit{population}, for a total of 10 equally-spaced repetitions across 10 seconds length. One, or more \textit{stimulation protocols} have been performed for each simulation by selecting a different target\textit{population}, with 1 second of separation between each \textit{stimulation protocol}. Each repetition's starting and ending times have been later used to compute the PSTH of the stimulated \textit{population}.
Different configurations have been tested with networks of 4 up to 60 \textit{populations}. Each simulation was generated and simulated with different seed values for the random number generator of our simulation, leading each time to a different set of both \textit{intra-} and \textit{inter-population} connections. Moreover, each simulation lasted for five minutes, divided into two sections: a \textit{background activity} section, lasting half of the recording, where only the noise and background components were active, and the \textit{stimulation} section, where stimulation protocols could be performed in addition to the basal components.
In particular, the background activity data was later used to train the RC model and to calculate the \textit{Intrinsic Connectivity Matrix}, while the stimulation data to perform response prediction tests.

\section*{Statistics}

For Fig. \ref{fig:connectivity general, algorithms}, a Kruskal-Wallis H-test has been performed to calculate if the population median of all groups is equal considering each metric. Subsequently, a series of two-sided Mann-Whitney-Wilcoxon tests with Bonferroni correction has been used to compare pairwise the algorithms' performance.

\section*{Code details}
All simulations were performed using the NEST simulator (version 3.5) \cite{Gewaltig:NEST} with custom code implemented in Python (version 3.8.17), with a kernel resolution of 0.01 ms. The RC model has been developed in Python implementing new classes on top of the ReservoirPy library (version 0.3.9) \cite{Trouvain2020}. Common Python libraries, such as Numpy, Scipy, and Scikit-learn have been used throughout the custom code for analysis. SpiCoDyn version 3.3 \cite{pastore2018s} has been used to perform Cross-Correlogram (CC) and Transfer Entropy (TE) analysis. 

\section*{Code availability}

The code for simulations, the RC model, and data analyses in this work will be available in a public repository. The code will be shared upon request.

\section*{Data availability}

The data of the simulations, recordings, and data analyses reported in this study will be available in a public repository. Data can be shared upon request.

\section*{Acknowledgements}
This work has received funding from European Union’s Horizon 2020
research and innovation program under the Marie Sklodowska-Curie grant
agreement No 101033260 (project ISLAND) and the European Research
Council (ERC) grant agreement No 788793 (project BACKUP). We also acknowledge Fondazione Caritro for their additional funding.

We would also like to thank Prof. Paolo Bettotti (University of Trento) for his support in simulations and computing infrastructures, Dr. Beatrice Vignoli (University of Trento) for her expert advice in the biological settings, Prof. Michela Chiappalone (University of Genoa) for the training in MEA measurements and data analysis.

\clearpage
\section*{Supplementary Figures}
\renewcommand\thefigure{Supplementary \arabic{figure}}
\setcounter{figure}{0}

\begin{figure}[h!]
    \centering
    \includegraphics[width = 0.75\textwidth]{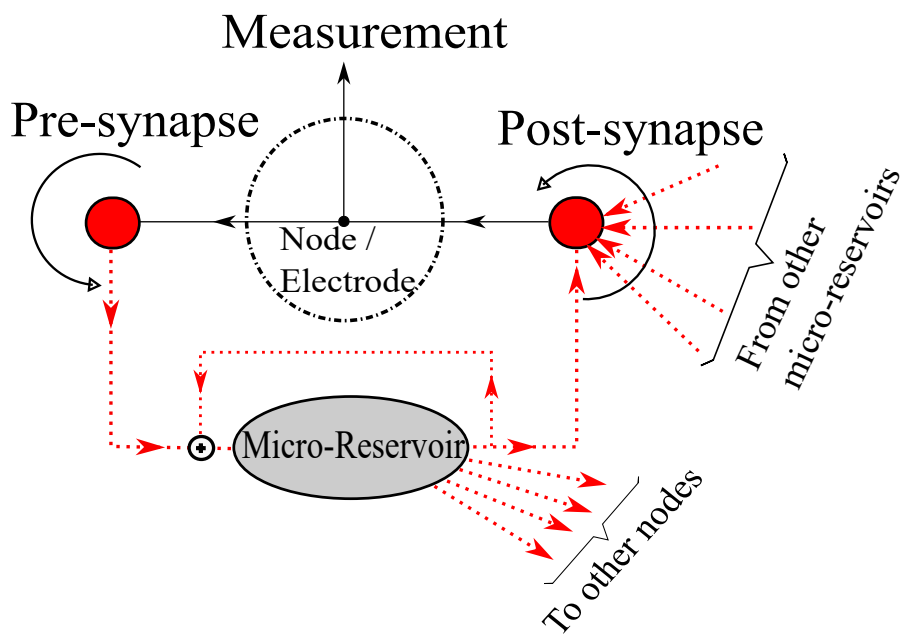}
    \caption{Enhanced visual representation illustrating the interconnection between the computational circuitry of the RC model (Fig. \ref{fig: artificial neural network}) and the intricate circuitry of biological signals.}
    \label{fig: Synaptic description}
\end{figure}

\begin{figure}[h!]
    \centering
    \includegraphics[width=0.75\linewidth]{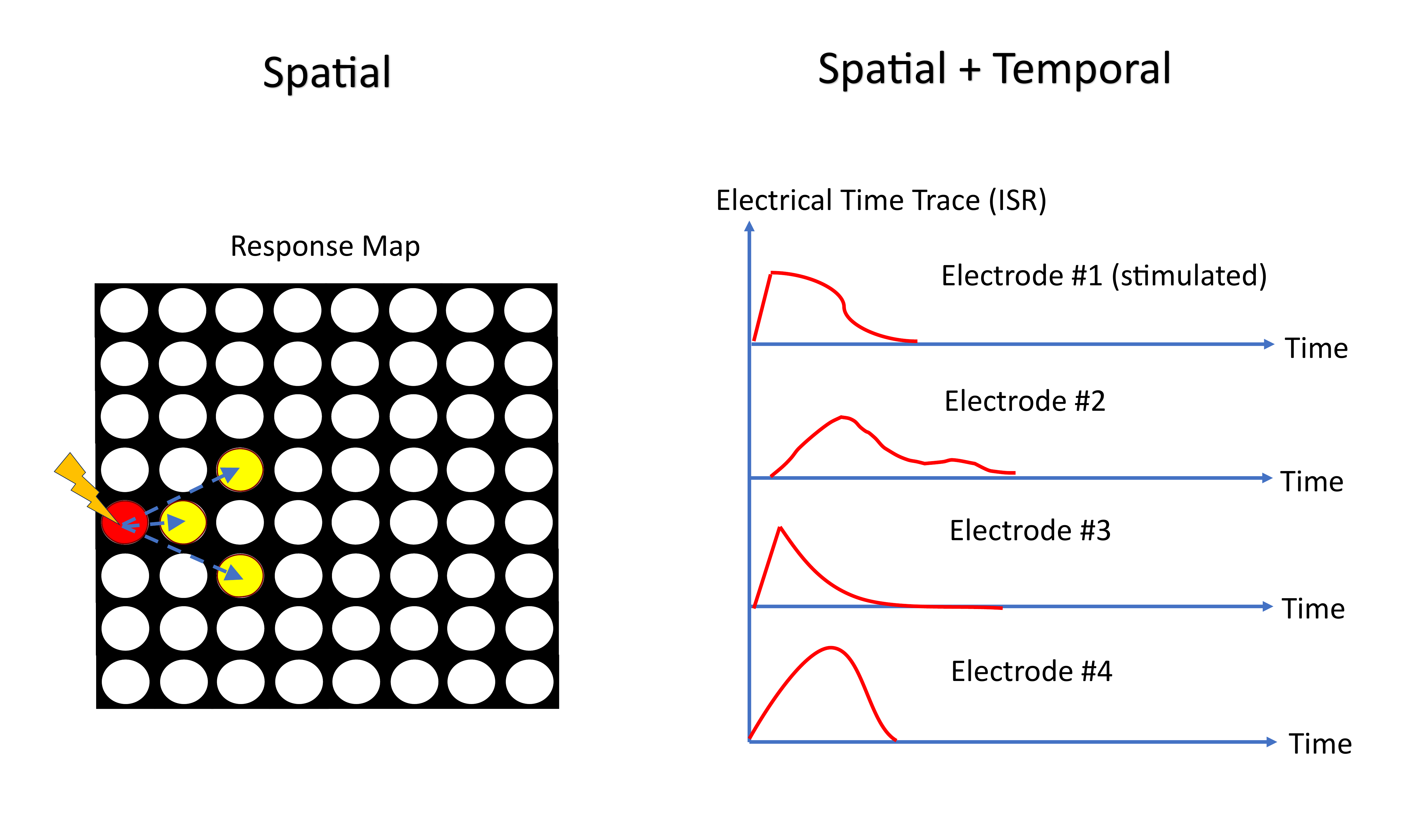}
    \caption{Illustration of the stimulus test designed to assess the RC model's network response prediction performance to a local stimulus. On the left, an electrode map highlights the stimulated electrode (in red) and the responsive electrodes (in yellow), demonstrating spatial prediction. This is further evaluated against the experimental/simulated response using the ROC AUC metric. On the right, a detailed temporal response from all involved electrodes showcases spatio-temporal prediction. This is quantitatively compared to the experimental/simulation test using the $\Bar{R}$ metric (refer to Eq. \eqref{eq: R bar}).}
    \label{fig:Stim1}
\end{figure}

\begin{figure}
    \centering
    \includegraphics[width=\textwidth]{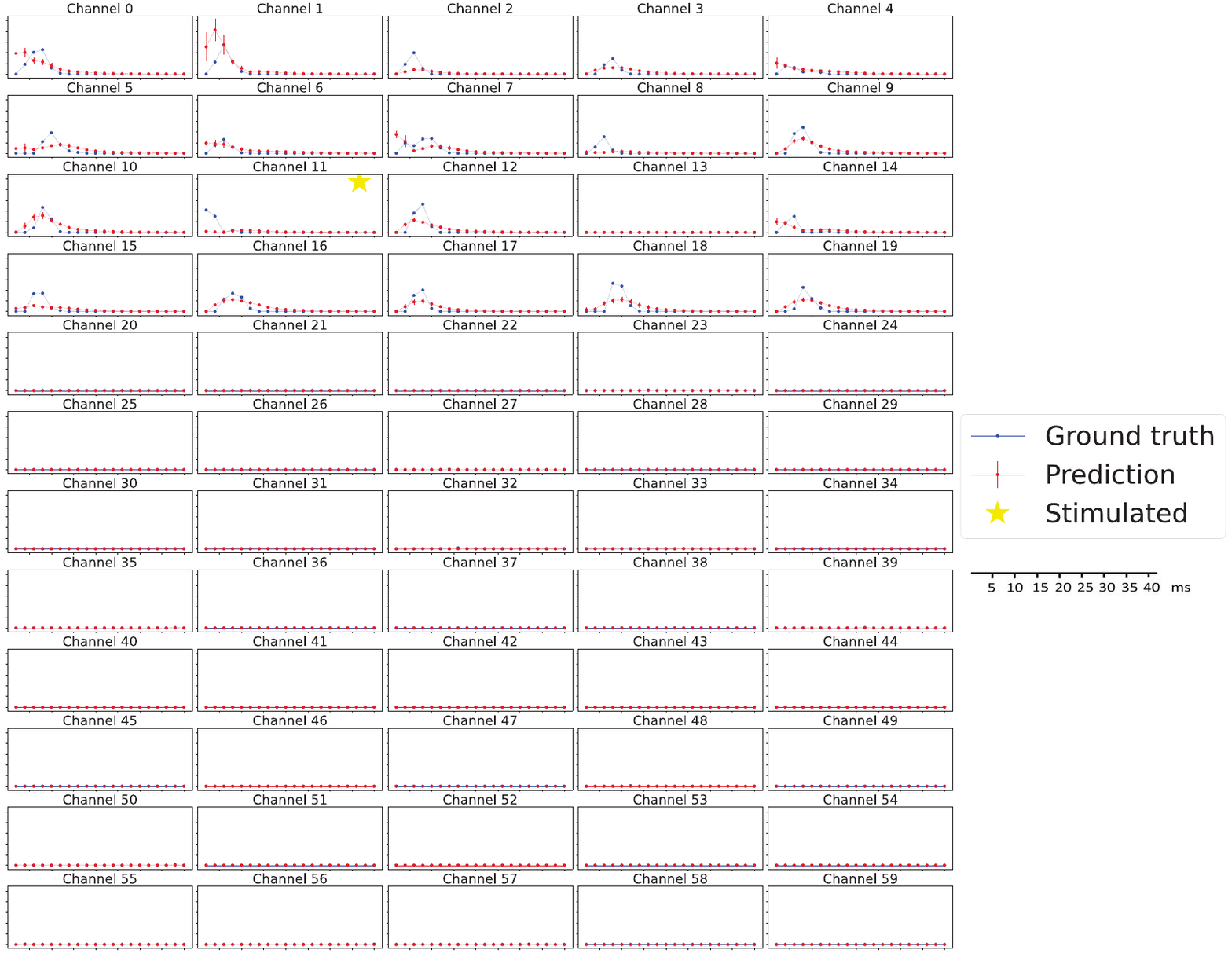}
    \caption{Response test (NEST simulation data). A detailed map of each channel's response to a stimulus is given at channel 11 (designated in yellow star). The time scale (X-axis) is given in the legend. Each point corresponds to a time step of 2ms. The $y$ axis is the normalized ISR in arbitrary units, scaled identically. $\Bar{R} = 0.158$, $\Bar{\tau}_l = 0.55ms$ at $\alpha = 0.5$ and $m=50$.}
    \label{fig:channel predictions sim}
\end{figure}
\begin{figure}
    \centering
    \begin{subfigure}[b]{0.33\textwidth}
        \includegraphics[width=\textwidth]{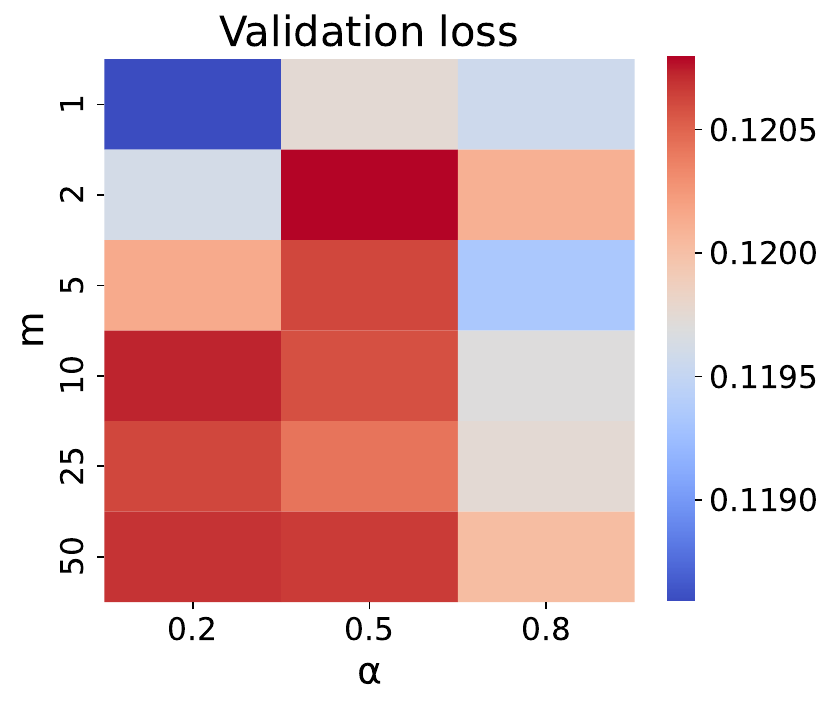}
        \caption{}
        \label{fig: m,alpha, validation}
    \end{subfigure}
    \begin{subfigure}[b]{0.33\textwidth}
        \includegraphics[width=\textwidth]{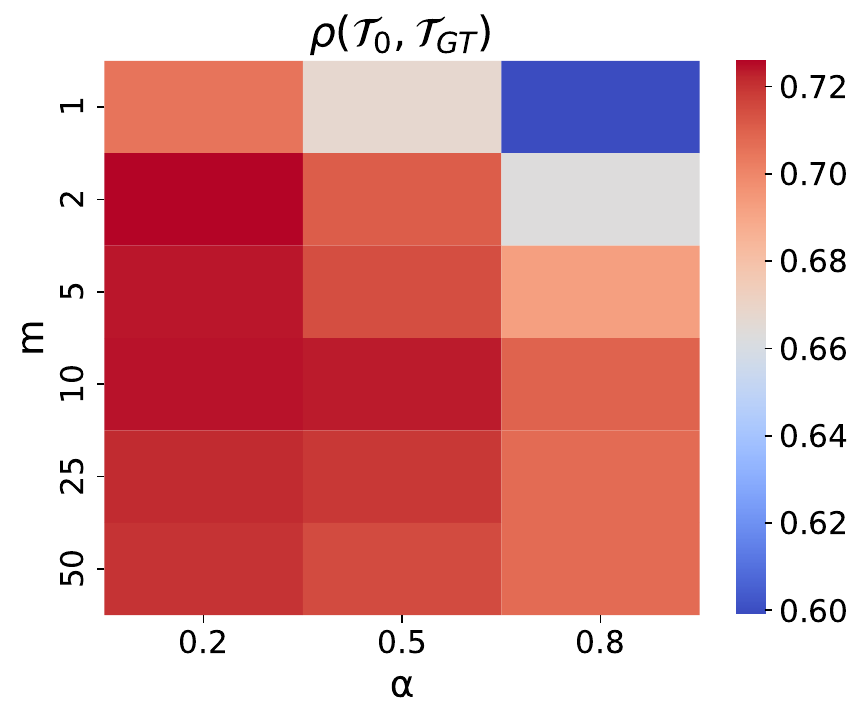}
        \caption{}
        \label{fig: m,alpha, rho_conn}
    \end{subfigure}
    \begin{subfigure}[b]{0.33\textwidth}
        \includegraphics[width=\textwidth]{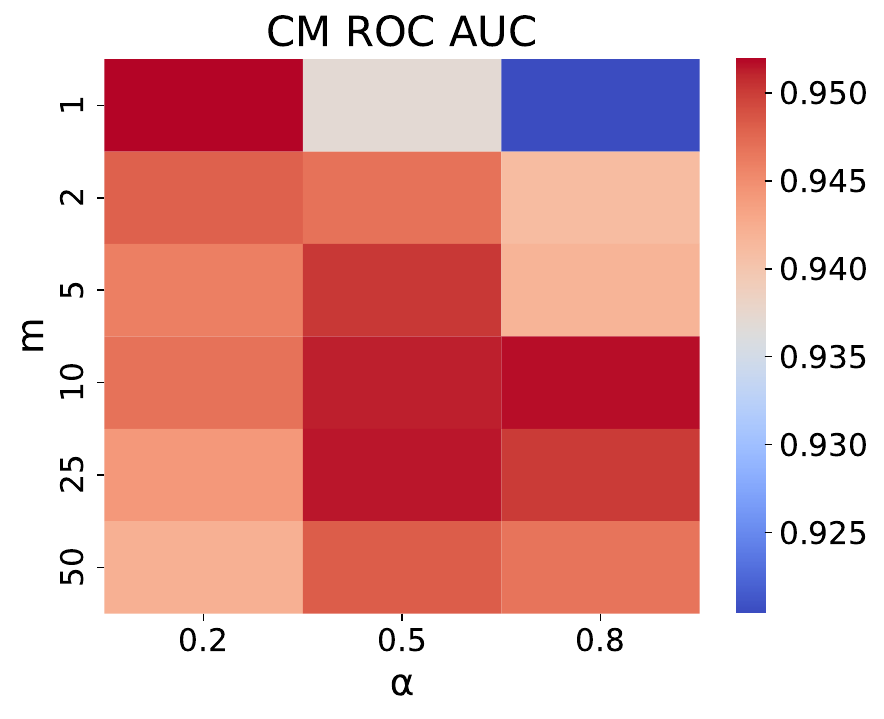}
        \caption{}
        \label{fig: m,alpha, auc_conn}
    \end{subfigure}
    \begin{subfigure}[b]{0.33\textwidth}
        \includegraphics[width=\textwidth]{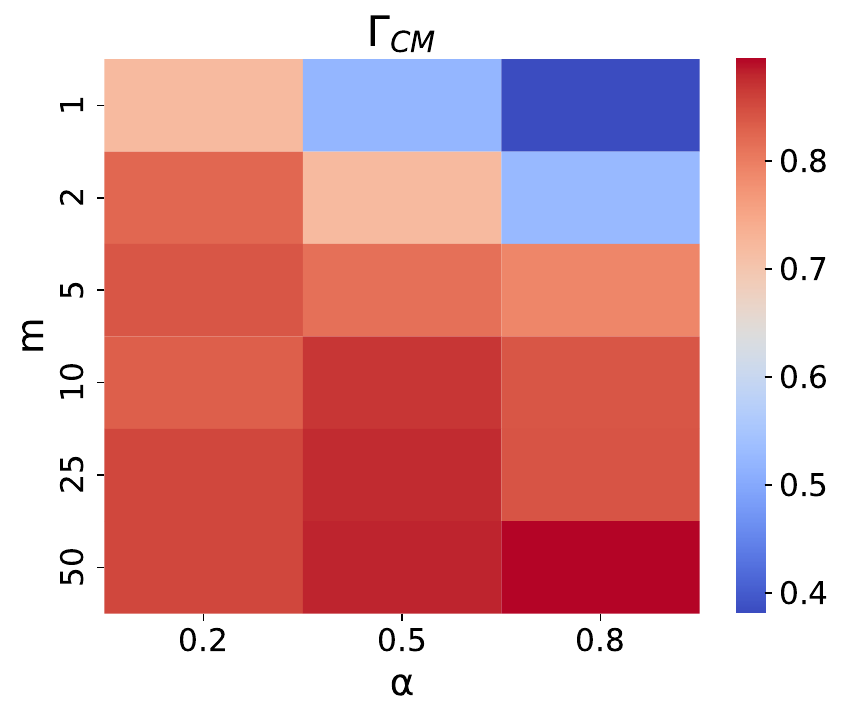}
        \caption{}
        \label{fig: m,alpha, conf}
    \end{subfigure}
    \begin{subfigure}[b]{0.33\textwidth}
        \includegraphics[width=\textwidth]{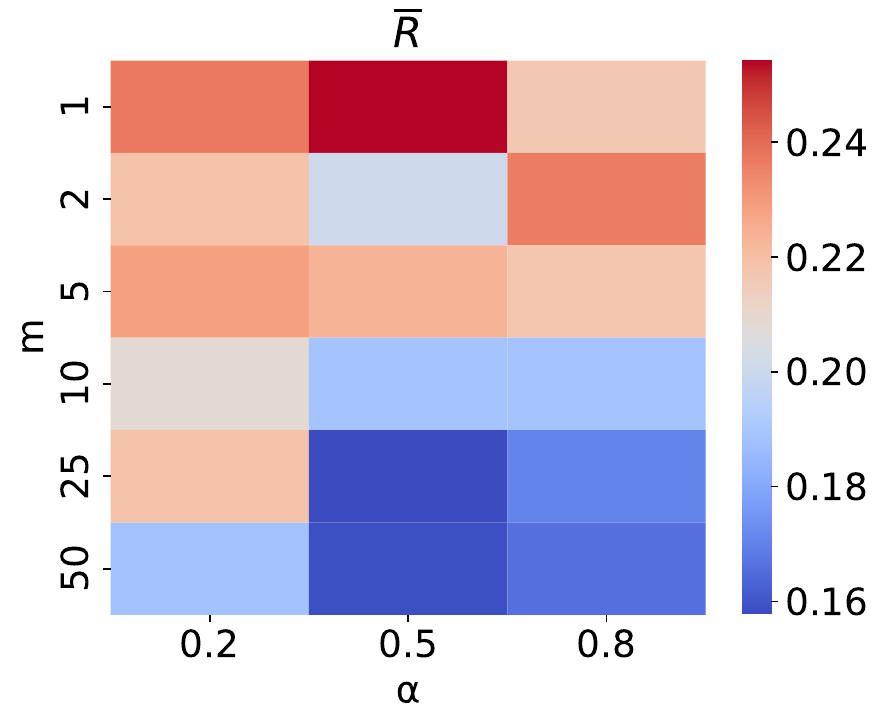}
        \caption{}
        \label{fig: m,alpha, R_bar}
    \end{subfigure}
    \begin{subfigure}[b]{0.33\textwidth}
        \includegraphics[width=\textwidth]{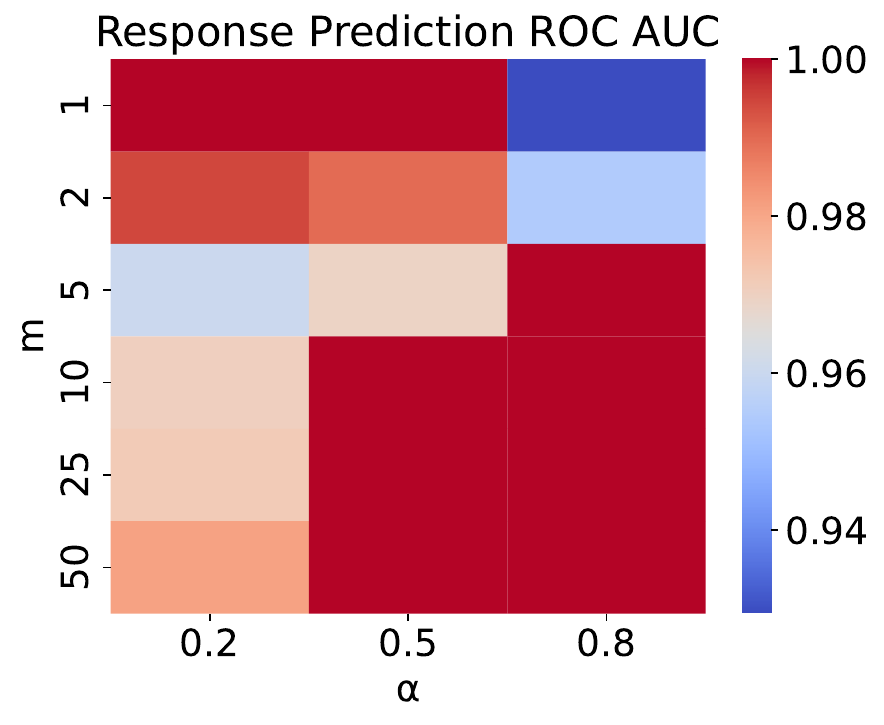}
        \caption{}
        \label{fig: m,alpha, auc_stim}
    \end{subfigure}
    \caption{Parametrization of RC model estimators and metrics (dependence on hyperparameters $m$ and $\alpha$). All figures refer to results obtained from the same NEST simulation dataset presented in Figs \ref{fig: CM example} and \ref{fig: response test example simulation}. (a) validation loss (Eq. \eqref{eq: Validation loss}). (b) Pearson correlation of the ICM with the ground-truth connectivity matrix. (c) ROC AUC for the connectivity prediction. (d) Confidence measure of the ICM (Eq. \eqref{eq: confidence}). (e) $\Bar{R}$ metric of the response prediction. (f) ROC AUC metric for the response prediction.} 
    \label{fig: m and alpha parametrization}
\end{figure}

\begin{figure}
    \begin{subfigure}{0.51\textwidth}
        \includegraphics[width=1\linewidth]{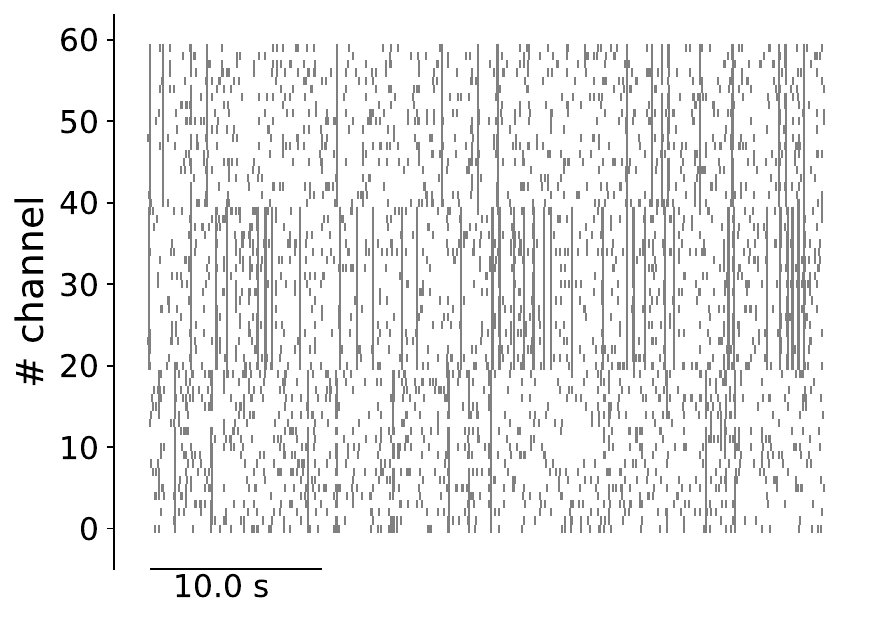}
        \caption{}
        \label{fig: raster}
    \end{subfigure}
    \begin{subfigure}{0.49\textwidth}
        \includegraphics[width=1\linewidth]{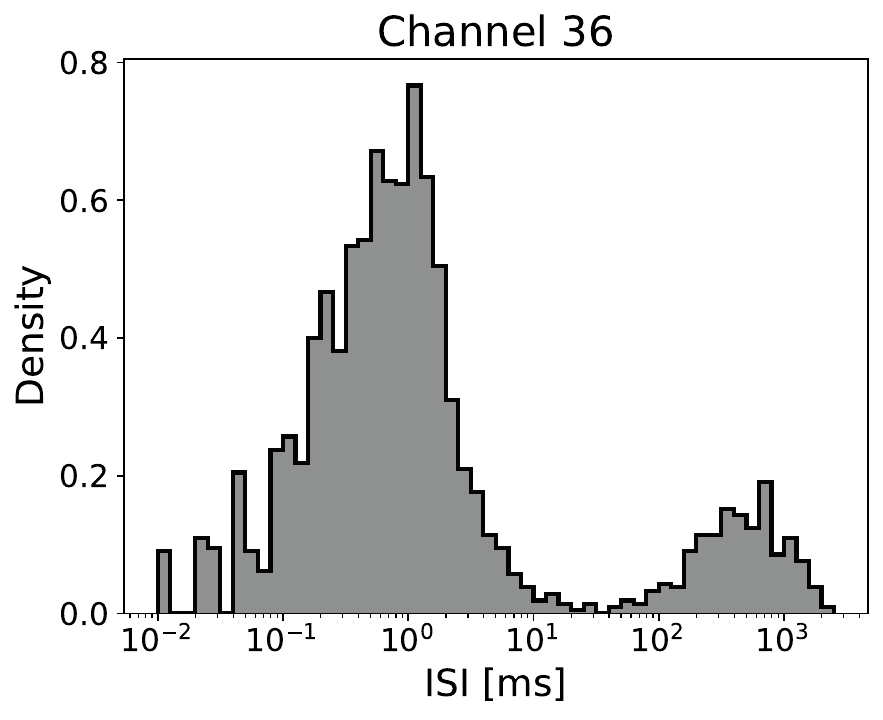}
        \caption{}
        \label{fig: ISI one channel}
    \end{subfigure}
    \begin{subfigure}{0.5\textwidth}
        \includegraphics[width=1\linewidth]{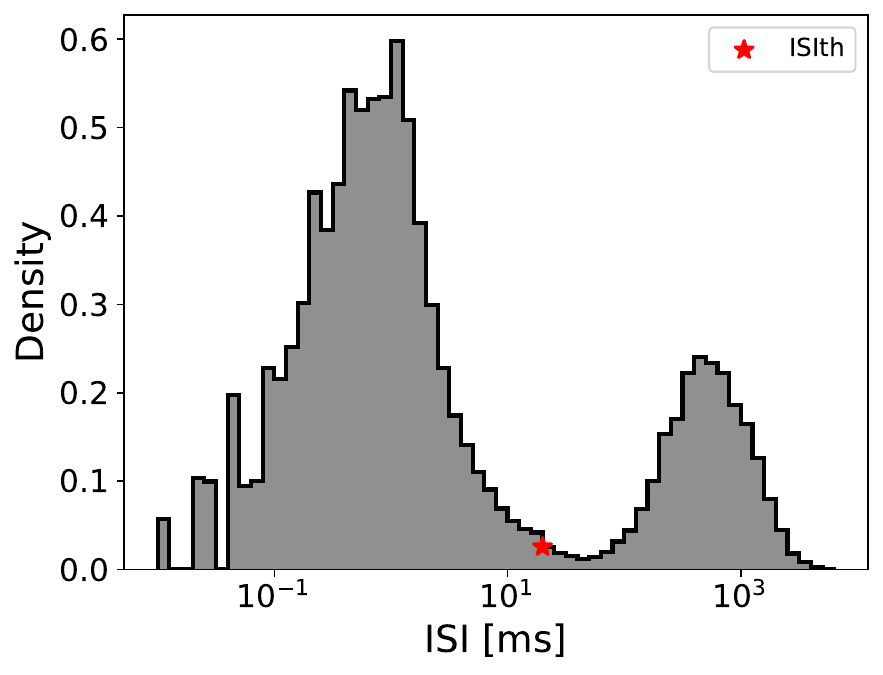}
        \caption{}
        \label{fig: ISI total}
    \end{subfigure}
    \begin{subfigure}{0.5\textwidth}
        \includegraphics[width=1\linewidth]{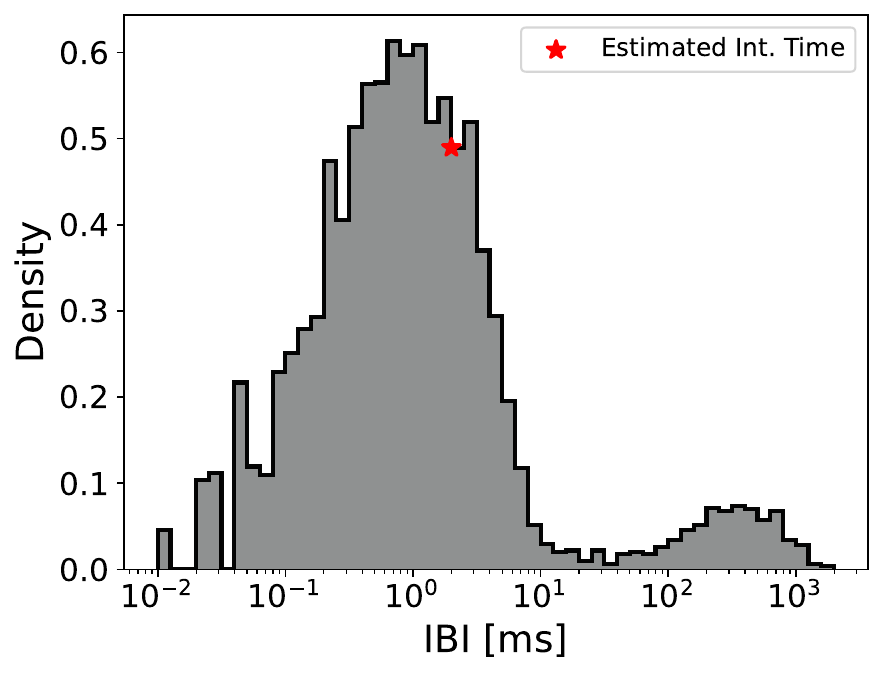}
        \caption{}
        \label{fig: IBI}
    \end{subfigure}
    \caption{Different network dynamics derived from electrophysiological data (taken from NEST simulation). (a) Raster plot illustrating the simulated electrophysiological activity of a 60-population \textit{in-silico} network. (b) An example of Inter-Spike-Interval (ISI) histogram from a single population, recorded on one channel, revealing a bimodal shape indicative of bursting activity, with a prominent peak at shorter time scales ($10^0 - 10^2$ ms). (c) Aggregated ISI histogram representing the entire network's ISIs across all channels (of intervals taken on a single channel), exhibiting a typical bimodal shape with a major peak in the $10^{-1} - 10^1$ ms range and a secondary peak in the $10^2-10^3$ ms range, signifying modulated bursting activity. Intra-burst intervals are characterized by the left peak, while inter-burst intervals are marked by the right peak. The red star indicates the ISI threshold ($ISI_{th}$) determined using the burst detection algorithm in Sec. \ref{sec: Spike and Burst Detection}. (d) Inter-Burst-Intervals (IBI) histogram showcasing the distribution of IBIs between different channels, depicting typical network bursting (NB) behavior. This histogram aids in deriving training data, divided into batches of NBs (Sec. \ref{sec: data structure}). The red star denotes the estimated integration time (in this case 2 ms), by the algorithm described in Sec. \ref{sec: integration time}.}
    \label{fig: Raster and Histograms}
\end{figure}

\begin{figure}
    \centering
    \begin{subfigure}[b]{0.33\textwidth}
        \includegraphics[width=0.95\textwidth]{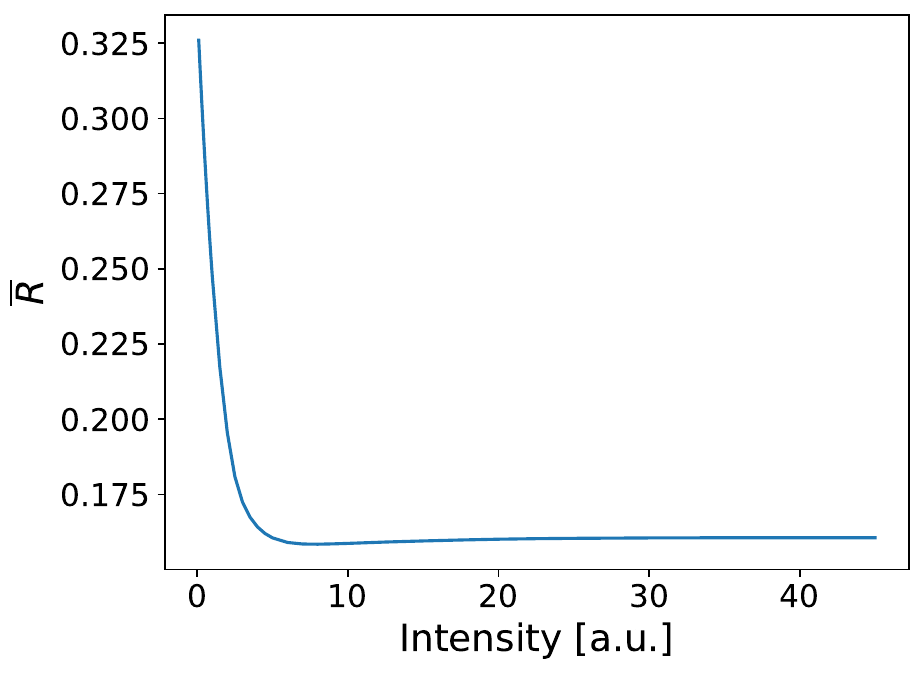}
        \caption{}
        \label{fig: R_bar vs intensity}
    \end{subfigure}
    \begin{subfigure}[b]{0.33\textwidth}
        \includegraphics[width=0.95\textwidth]{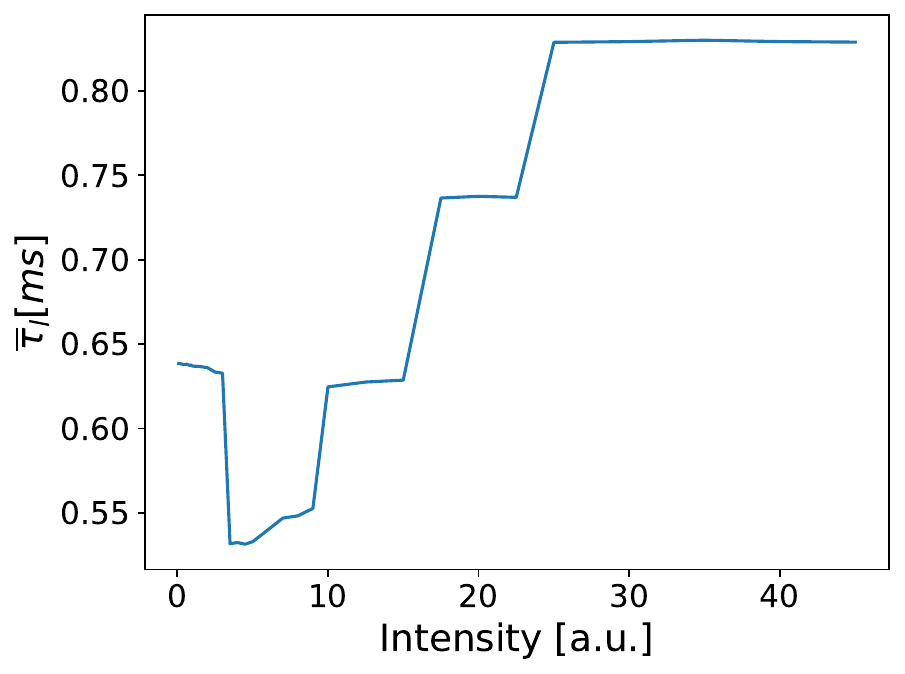}
        \caption{}
        \label{fig: tau vs intensity}
    \end{subfigure}
    \begin{subfigure}[b]{0.33\textwidth}
        \includegraphics[width=0.95\textwidth]{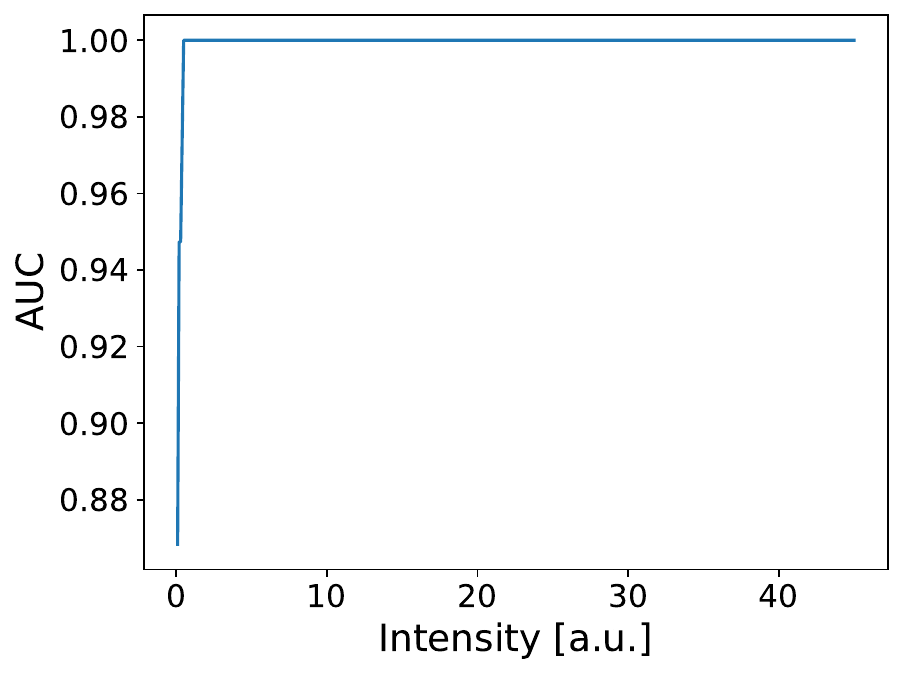}
        \caption{}
        \label{fig: auc vs intensity}
    \end{subfigure}
    \caption{Response test metrics dependence on input intensity of the stimulus. The intensity was used as a tuning parameter to fit the model response to the experiment. (a) $\Bar{R}$ (b) The overall time lag between the modeled response and the experiment, $\Bar{\tau}_l$ (Eq. \eqref{eq: tau_l}). (c) ROC AUC.}
\end{figure}
\begin{figure}
    \centering
    \begin{subfigure}[b]{0.5\textwidth}
        \centering
        \includegraphics[width=\textwidth]{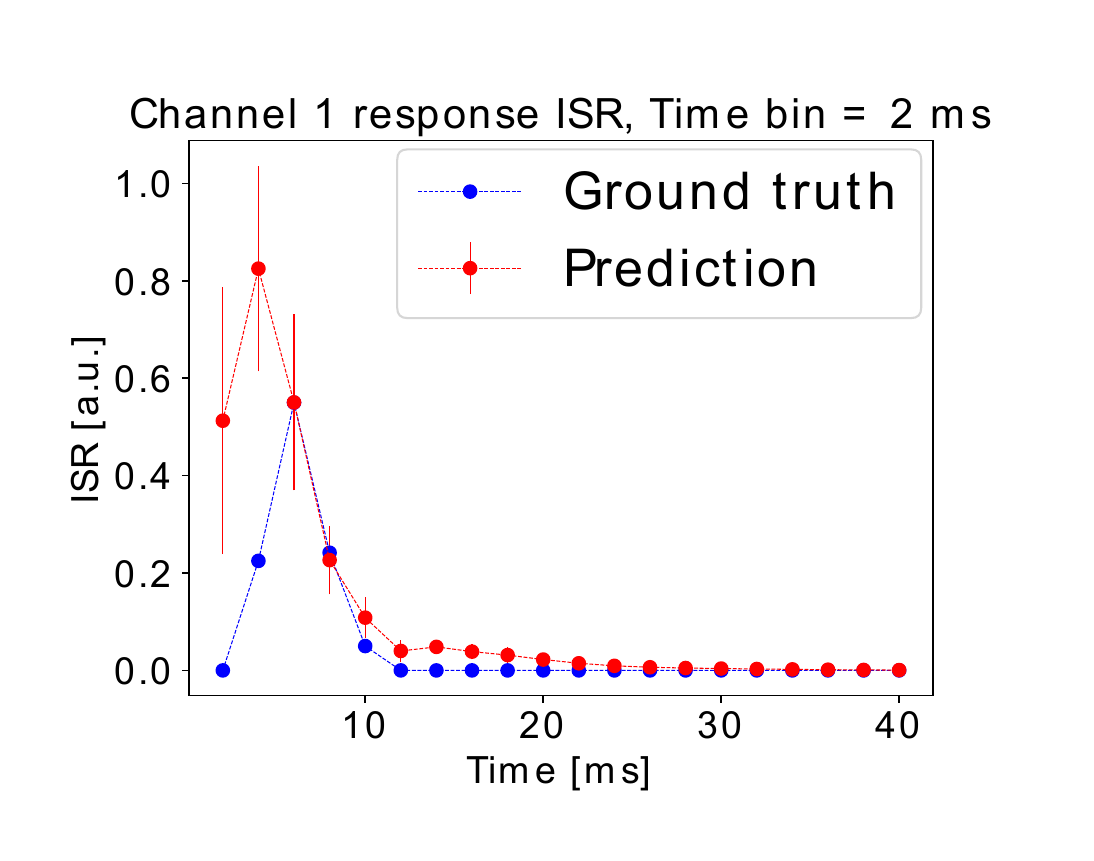}
        \caption{}
        \label{fig: channel prediction} 
    \end{subfigure}
    \begin{subfigure}[b]{0.48\textwidth}
        \centering
        \includegraphics[width=0.95\textwidth]{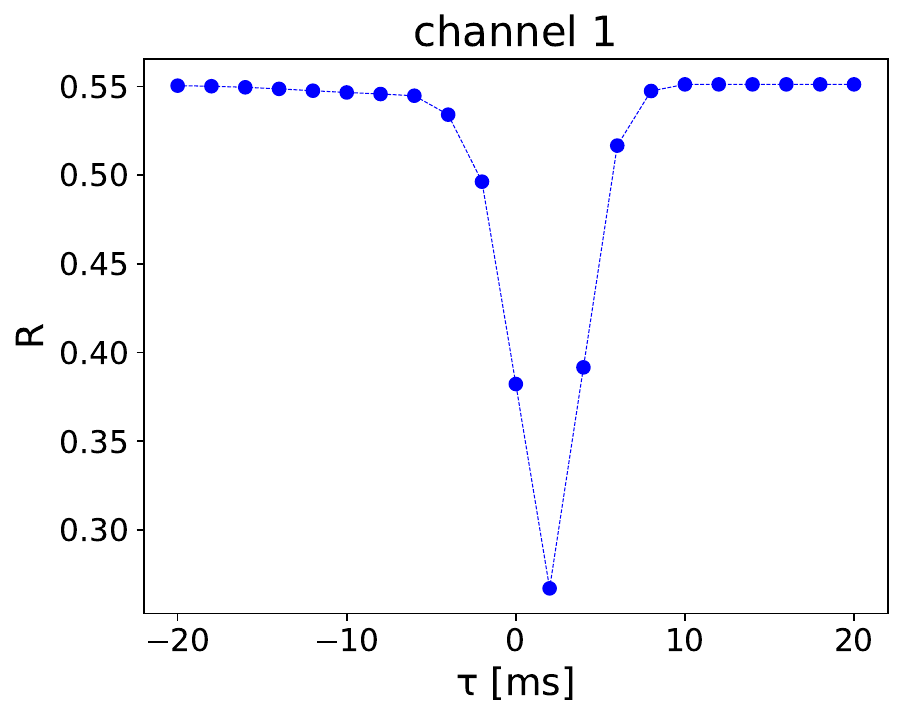}
        \caption{}
        \label{fig:} 
    \end{subfigure}
    \caption{The principle of the cross-root-mean- square-error ($R$) metric. (a) ground-truth and prediction data. (b) The outcome of $R$ for the prediction of (a)- Eq. \eqref{eq: xrmse}. The minima describes the channel error and the time lag ($\varepsilon_i$, $\tau_{l,i}$ Eq. \eqref{eq: channel error}) are obtained. This procedure is done over all the nodes of the network, followed by the total error, $\Bar{R}$, and time lag, $\Bar{\tau}_l$, estimation (Eqs. \eqref{eq: R_bar_detailed}, \eqref{eq: tau_l}).} 
\end{figure}

\clearpage
\section*{Supplementary Materials}

\setcounter{subsection}{0}
\renewcommand{\thesubsection}{\Roman{subsection}}

\subsection{Experimental procedures}

\setcounter{subsubsection}{0}
\renewcommand{\thesubsubsection}{\Roman{subsection}.\alph{subsubsection}}

\subsubsection{Culture Preparation}
In the preparation of primary neuron cultures, animals at 17/18 days of gestation were utilized, and all procedures conducted at the University of Trento in Italy strictly adhered to the approved ethical guidelines and regulations. To extract neurons from embryonic cortex tissue, we used the following protocol: initially, we decapitated the embryos and dissected their cortex under sterile conditions within a laminar flow cabinet, utilizing a standard dissection buffer enriched with glucose (HBSS). Afterward, we replaced the dissection buffer with 5ml of 0.25\% trypsin-EDTA (Gibco) and allowed the tissues to incubate for 20 minutes to promote cell dissociation. Subsequently, to halt the action of trypsin, we introduced 5ml of DMEM medium supplemented with 10\% FBS. We gently pipetted the solution and employed an appropriate strainer for neuron isolation. Following this, we subjected the separated cells to centrifugation at 1900 rpm for 5 minutes, effectively removing the superficial solution (DMEM,10\% FBS, P/S). Finally, we added a seeding medium comprising DMEM, 10\% FBS, and P/S until the cell concentration reached 1700 cells/$\mu$l. In our experimental procedure, we carefully dispensed 80$\mu$l of cell solution into each chip, employing a droplet technique and targeting the central region of the chips. These chips had been pre-coated with Poly-D-Lysine (PDL) and laminin. After a 2.5-hour incubation period, we proceeded to introduce the nourishing medium, consisting of Neurobasal supplemented with 1\% B27, 1\% Glutamax, and 1\% P/S.

To ectopically express ChR2 in primary neuronal culture, we diluted 0.2$\mu$l of pAAV-hSyn-hChR2(H134R)-EYFP (\#26973) in 100$\mu$l of standard feeding medium, which is composed of Penicillin-Streptomycin (10,000 U/mL) (1\% v/v), GlutaMAX Supplement (1\% v/v), B-27 supplement (50X) (2\% v/v) and Neurobasal. All of these substances were purchased from Gibco. We then incubate the culture for 24 hours with the previously mentioned viral vector solution. Half of the medium was replaced with a standard feeding medium following the incubation period. After six or seven days, the expression of EYFP can be used to track the expression of ChR2.

\subsubsection{MEA Recordings}
    The electrophysiological signals were recorded using the MEA-2100mini system of \textit{Multichannel Systems} GmbH (MCS). The microelectrode array chips used in our experiments were 60MEA-200/30iR-Ti-gr by MCS, which are chips with 60 titanium-nitride electrodes embedded in glass and surrounded by a glass ring. The electrodes are of 30$\mu$m diameter, where the horizontal and vertical spacing between each pair of electrodes is of 200$\mu$m. The MEA-2100mini system collects the signals through a head-stage device. Then the signals undergo amplification and filtering. The system is then connected to a PC through an interface board. The recording is performed using MCS experimenter software, where the signals can be digitally filtered, inceptively-analyzed, and tracked in real time. We sampled the signals at 20KHz. The recorded files are then saved and exported for a secondary offline analysis.

\subsubsection{Electrical Stimulation}

The MEA-2100 mini system has also the feature of providing electrical stimulation to neuronal cultures. In our experiments, we applied bi-phasic square-shaped pulses with a peak-to-peak amplitude of 1.6V and a duration of 40 microseconds (20 microseconds for each phase) at a frequency of 0.5Hz for 300 repetitions.

\subsubsection{Optical Stimulation}
        
    In some experiments, we employed optical stimulation to activate ChR2-expressing neurons. This method enables more precise and localized manipulation of neural activity, stimulating specific regions within the network, as opposed to the broader influence of electrical stimulation. As the light source, we used a Digital Light Processor (DLP) system. The system is a DLP E4500, which includes 3 LEDs, optics, a WXGA DMD (Wide Extended Graphics Array Digital Micro-mirror Device), and a driver board. The blue LED (488nm) which is used in this work has a power of 600mW. The light from the LEDs impinges on the DMD which has 1039680 mirrors arranged in 912 columns by 1140 rows with a diamond array configuration. This system allows patterned illumination with pre-loaded and custom patterns that can be chosen through the DLP E4500 software. Moreover, these patterns can be pulsed in time, with both an internal or external trigger, with a nominal precision down to $\mu$s. The system supports 1-, 2-, 3-, 4-, 5-, 6-, 7-, and 8-bit pixel resolution images with a 912 columns $\times$ 1140 rows resolution, meaning that each pixel corresponds to a micro-mirror on the DMD. The light coming from the DLP system is collimated and aligned to the optical path of the microscope from the rear port of the system. It is then collected by a macro TAMRON 90mm AF2.5 objective and the light pattern is imaged on the sample plane, where the MEA chip is located, a by 10$\times$ objective, while passing through a dichroic mirror (Chroma T505lpxr-UF1) which acts like a high-pass filter, reflecting all the wavelengths smaller than 505nm.
        
    The ChR2-infected culture could be imaged using a microscopy system, where the signal of the green fluorescent protein (E-YFP) expressed by the ChR2-infected cells, is transmitted through a dichroic filter and detected by a CMOS camera. The culture image allows to direct the desired light pattern from the DLP directly to the region of interest in the culture.
    
    In this work, the light stimulation comprised 300 pulses of 20 milliseconds each, delivered as blue light at a frequency of 0.5 Hz (with 2-second intervals between pulses.) Specifically, we stimulated an area under a light spot with a diameter of around 50$\mu$m, around a target electrode and recorded the electrical response from all the network by MEA. 

\subsection{Data Pre-Processing}
This sub-section details the electrophysiological data preprocessing method for the training datasets employed in the RC model. This process converts the raw electrical signals into data batches that capture episodes of notable culture activity. It's worth noting that the intermediary algorithms within this procedure are flexible and can be tailored to specific requirements since they are not essential for the model's functionality.

\subsubsection{Spike and Burst Detection}
\label{sec: Spike and Burst Detection}
    Following MEA recordings, the raw data was exported from MCS software and analyzed using a custom code written in Python. First, the raw signals recorded on MEA were digitally filtered with a band-pass Butterworth filter with cutoff frequencies of 0.3 and 3 KHz. Spikes were detected using the PTSD algorithm \cite{maccione2009novel}, setting the differential threshold (DF) to 8, the refractory period (RP) to 1 ms, and the peak lifetime period (PLP) to 2 ms.

    After detecting spikes, we proceeded to apply a burst detection algorithm inspired by the approach outlined in \cite{pasquale2010self}, but with some minor modifications. By utilizing the spike trains, we derived the inter-spike intervals (ISI) for each channel within the recording. Subsequently, data from all channels was aggregated into a single ISI histogram (Fig. \ref{fig: ISI total}), constructed with fixed bins of $\Delta\log_{10}\{ISI\} = 0.1$ (in units of $\log_{10}$(ms). The ISI threshold was then calculated by the following algorithm:
    \begin{itemize}
        \item Detect local maxima in the ISI histogram.
        \item Sort maxima by their significance, where significance is determined by $S_{peak} = p\cdot w$; $p$ is the peak prominence and $w$ is the peak full-width at half-prominence (FWHP), expressed in $\log_{10}(ms)$ units.
        \item Choose the most significant peak within the bins between $1$ and $10$ ms, characterizing typical inter-burst intervals \cite{pasquale2010self}.
        \item Select the most significant peak within the bins spanning from $1$ to $10$ ms. This peak characterizes the typical fast \textit{intra-burst intervals} (for each channel).
        \item Choose the most significant peak within the bins greater than $10$ ms. This peak characterizes the typical \textit{inter-burst intervals} (for each channel).
        \item calculate the ISI threshold (in ms) by:

    \begin{equation}
        \log_{10}(ISI_{th})= \frac{x_l+w_l/2+x_r-w_r/2}{2}
    \end{equation}

    where $x_l,x_r$ are the bin value of the left and right peaks, respectively; $w_l,w_r$ are the FWHP of the left and right peaks, respectively- all expressed in $\log_{10}(ms)$ units.
    \end{itemize}
Following that, bursts were identified in each temporal sequence, provided that it contained a minimum of three spikes with inter-spike intervals $ISI^{(i,j)} \leq ISI_{th}$, where $i,j$ are the burst and channel indices, respectively. Consequently, bursts starting time $t_B^{i,j}$ were registered.

In the subsequent phase, we examined instances where bursting activity spread across the culture, indicating periods when bursts were detected on multiple channels within a specific time frame - also known as \textit{Network Bursts} (NBs) \cite{pasquale2010self,van2004longterm,chiappalone2005burst}. NB was identified if bursts occurred on at least two distinct channels $j$ and $k$ with respective starting times, $t_{B}^j$ and $t_{B}^k$, were separated by a time interval no greater than
$\langle L_B \rangle /2$, where $\langle L_B \rangle$ is  the average length of a burst. Consequently, NB starting times $t_{NB}^{j}$ were registered, followed by the procedure described in Sec. \ref{sec: Training}.

\subsubsection{Estimation of the Integration Time}
\label{sec: integration time}

We define integration time as the duration required for each channel (neuronal circuit) to assimilate incoming information and transmit a response through the network. A detailed analysis of bursting dynamics, represented by the Inter-Burst-Interval (IBI) histogram showcasing the distribution of IBIs between distinct channels (refer to Fig. \ref{fig: IBI}), reveals a prominent peak. This peak represents a characteristic range of time intervals during which one channel responds to the burst of another. Hence, we deduce that information flows within this specific rate, leading us to designate it as the effective integration time.

Consequently, we utilize this characteristic time to sample the state $\mathbf{y}$ of the network at time steps corresponding to the integration time. Subsequently, we bin the spike trains into intervals aligning with this time, generating the Instantaneous Spike Rate (ISR) traces (see Sec. \ref{sec: data structure}).

Accordingly, we implemented the following algorithm for estimating the integration time:
\begin{itemize}
    \item Detect local maxima in the IBI histogram.
    \item Sort maxima by their significance, where significance is determined by $S_{peak} = p\cdot w$; $p$ is the peak prominence and $w$ is the peak full-width at half-prominence (FWHP), expressed in $\log_{10}(ms)$ units.
    \item If the most significant peak is found at a value less than 2 ms $\rightarrow$ set 2 ms as the integration time. This assumption arises from the rapid dynamics, and typically, this interval can accommodate at most one or two spikes.
    \item If no peaks are found in this range $\rightarrow$ set 5 ms as the integration time. This pertains specifically to slow dynamics; unless the network's dynamics are rich within this regime, poor performance is expected.
    \item Otherwise $\rightarrow$ the most significant peak in this range indicates the integration time, with its value is rounded to the closest 0.1ms multiple (for usage of rounded numbers in the binning step).
\end{itemize}

Through empirical testing of the RC model's performance across various integration times, we demonstrated that the algorithm consistently delivers close-to-optimal results in a substantial number of cases.

\subsection{Model Parametrization}
\label{sec: parameters}
We can categorize the model's parameterization (Section \ref{sec: Methods}) into three distinct types: fixed parameters- matrices $\mathcal{W}_{res}$, $\mathcal{W}_{in}$, and $\hat{\mathcal{S}}$; trainable parameters represented by $\mathcal{W}_{out}$; and hyperparameters $m$ and $\alpha$.

As discussed in Section \ref{sec: Methods}, the fixed parameters consist of randomized matrices adhering to specific constraints. One of our key assessments of the model involved investigating the stability of the model's predictions (connectivity or response) concerning the initialization of these fixed matrices. To achieve this, we conducted the training process a total of $N_{rep}$ times (typically between 5 and 10 repetitions) and examined the consistency of the model's outcomes.

For instance, in the lasso regression (Eq. \ref{eq: lasso regression}), the optimal $\mathcal{W}_{out}$ was determined for each set of $\mathcal{W}_{res}$, $\mathcal{W}_{in}$, and $\hat{\mathcal{S}}$. We specifically evaluated the stability of connections within the intrinsic connectivity matrix (ICM), $\mathcal{T}_0$ (Eq. \ref{eq: intrinsic conn matrix}). This matrix results from the product of $\mathcal{W}_{out}$ and $\hat{\mathcal{S}}$ with $\mathcal{W}_{in}$. We assessed the consistency of connections in the ICM across different training sessions, quantifying it as the ICM confidence as follows:
\begin{equation}
    \label{eq: confidence}
    \Gamma_{CM} = 1 - \frac{\max(\sigma_{\mathcal{T}_0})}{\max|\langle \mathcal{T}_0 \rangle |}
\end{equation}
Here, $\Gamma_{CM}$ represents the confidence value, $\sigma_{\mathcal{T}_0}$ stands for the standard deviation of the connection weights across the repetitions, and $\langle \mathcal{T}_0 \rangle$ denotes the mean connection weights averaged over the repetitions.
The assessment of response prediction stability, as outlined in Section \ref{sec: response test}, involved conducting the test with multiple initializations and calculating the mean response. The degree of stability is visualized by the error bars, which represent the standard deviation of the predicted response for each channel  (Figs. \ref{fig:channel predictions sim} and \ref{fig:channel predictions exp}).

Regarding the hyperparameters $m$ and $\alpha$, we characterized their effect on model outcomes in all steps: Validation, Connectivity prediction, and Response prediction. Fig. \ref{fig: m and alpha parametrization} illustrates the characterization of some of the model's metrics as a function of hyperparameters $m$ and $\alpha$.

\subsection{External Test Parameters (Stimulation Intensity)}
\label{sec: intensity}
During the test phase where we examined how accurately the model can predict the network response to a given input (stimulus), we found that it is useful to tune the stimulation intensity as a fitting parameter to minimize the error between the ground-truth response $\mathbf{U}$ and the prediction $\Tilde{\mathbf{U}}$ in terms of $\Bar{R}$ metric (Eq. \eqref{eq: R bar}). Assuming that the stimulation is applied around node (electrode) $p$, we represent the input state (the stimulation) as:
\begin{equation}
    \mathbf{u}_{in} = I\cdot \delta_{i,p} \;\;\; i=1,2,...,N_{ch}
\end{equation}
where $I$ is the intensity and $\delta_{i,p} =0 $, for $i\neq p$ and $\delta_{i,p}=1 $ for $i=p$ (Kronecker delta).

\subsection{Performance Optimization}

The model underwent training with a specific set of hyperparameters $(\alpha, m)$, as detailed in Section \ref{sec: parameters}. Each training set was iterated five times, employing different initializations for the fixed matrices of the model. During the testing phase, we assessed the trained model for each hyperparameter set using $\Bar{R}$ and ROC AUC metrics (refer to Section \ref{sec: response test}). The primary optimization parameter was the stimulus intensity, outlined in Section \ref{sec: intensity}. The optimal intensity was determined based on minimizing $\Bar{R}$ for each combination of $\alpha$ and $m$. In instances where the $\Bar{R}(I)$ characteristic exhibited a flat trend, the ROC AUC metric was employed as the arbitrator.

Subsequently, we generated maps of $\Bar{R}(\alpha, m)$ and $AUC(\alpha, m)$ by averaging across various stimulation protocols. From these maps, we identified the optimal values for $m$ and $\alpha$. The optimal $m$ was regarded as a global parameter applicable to both stimulus testing and connectivity retrieval, as it is indicative of the network's overall complexity and structure. In contrast, the optimal $\alpha$ was chosen independently for the stimulus test and connectivity, recognizing its dependence on specific operating regimes.

The optimization methodology employed in the model presented in this work hinges on test metrics rather than relying solely on training and validation phases. Anticipated is the need for fine-tuning the optimization process to align with the demands of real-time applications. Consequently, it is imperative to conduct additional studies to delve deeper into and refine this approach.

\clearpage
\bibliography{paper}

\end{document}